\documentclass[letter,11pt,notitlepage,amssymb]{article}
\pdfoutput=1
\usepackage{graphicx}
\usepackage[margin=30mm]{geometry}
\usepackage{setspace}
\usepackage{amsmath}
\usepackage[font={small}, labelfont={bf,small}]{caption}
\usepackage{siunitx}
\usepackage{cite}

\usepackage{hyperref}

\usepackage[usenames,dvipsnames,svgnames,table]{xcolor}
\usepackage{latexsym}
\usepackage{url}

\usepackage{xspace}
\usepackage{amssymb}
\usepackage{amsmath}
\usepackage{bm}
\usepackage{booktabs}

\usepackage{setspace} 
\newcommand{\high}{}
\newcommand{\shigh}{}

\DeclareCaptionLabelSeparator{bar}{\,$\boldsymbol{\vert}$\,}
\newcommand{\FigComponentsEvolution}{\ref{fig:GC_CO}}
\newcommand{\FigHastagsNetwork}{\ref{fig:ht_cooc}}
\newcommand{\FigDailyOpinionSCGC}{\ref{fig:num_users_in_gc}}
\newcommand{\FigDailyOpinion}{\ref{fig:num_users}}

\newcommand{\FigNYTFitRMSER}{\ref{fig:fit_rmse_r}}
\newcommand{\FigPollPrediction}{\ref{fig:nyt_partial_fit}}
\newcommand{\FigUsersActivity}{\ref{fig:num_tweets_per_user_CCDF}}

\newcommand{\TabHashtagsExamples}{\ref{tab:hstg_lists_1st}}
\newcommand{\TabClassificationScores}{\ref{tab:cross_val_score}}
\newcommand{\FigCompBenchmark}{\ref{fig:fit_polls_wl13_benchmark}}

\newcommand{\FigFilteredResults}{\ref{fig:num_userskwrd_filt}}
\newcommand{\FigRobustnessCheck}{\ref{fig:num_users_random}}
\newcommand{\FigFinalvsInitial}{\ref{fig:fit_pears_rmse_initial_vs_final}}
\newcommand{\FigHTNetworkClimateChange}{\ref{fig:ht_cooc_signi_climate_change_final}}
\newcommand{\FigHTNetworkFrench}{\ref{fig:ht_cooc_signi_french_elections_final}}
\newcommand{\TabHTjunsep}{\ref{tab:hashtags_final_set}}
\newcommand{\TabHTsepnov}{\ref{tab:hashtags_final_set_sep_nov}}

\begin{document}

 \begin{center}

 \LARGE{\textbf{Validation of Twitter opinion trends with national polling 
aggregates: Hillary Clinton vs Donald 
Trump}}
 
\vspace{1cm}

\large Alexandre Bovet, Flaviano Morone, Hern\'an A. Makse

\vspace{0.2cm} \normalsize \textit{Levich Institute and Physics
  Department, City College of New York, New York, New York 10031, USA}

\end{center}

\begin{abstract}

Measuring and forecasting opinion trends from real-time social media
is a long-standing goal of big-data analytics.
Despite its importance, there has been no conclusive scientific evidence so far
that social media activity can capture the opinion of the general
population.  Here we develop a method to infer the opinion of Twitter
users regarding the candidates of the 2016 US Presidential Election by
using a combination of statistical physics of complex networks and
machine learning based on hashtags co-occurrence to develop an
in-domain training set approaching 1 million tweets.
We investigate the social networks formed
by the interactions among millions of Twitter users and infer the
support of each user to the presidential candidates.  The resulting
Twitter trends follow the New York Times National Polling Average,
which represents an aggregate of hundreds of independent traditional
polls, with remarkable accuracy.
Moreover, the Twitter opinion trend precedes the aggregated NYT polls by 
10 days, showing that Twitter can be an early signal of global
opinion trends.
Our analytics unleash the power
of Twitter to uncover social trends from elections, brands to
political movements, and at a fraction of the cost of national polls.

\end{abstract}

Several works have showed the potential of online social media, in
particular of the microblogging platform Twitter, for analyzing the
public sentiment in general
\cite{mislove2010pulse,Hannak2012,Pak2010,Quattrociocchi2014} or to
predict stock markets movements or sales performance
\cite{Liu2007,Bollen2011,Zheludev2014,Ranco2015,Curme2015}.
{\high 
\label{intro}
With the increasing importance of Twitter in
  political discussions, a considerable number of studies
  \cite{OConnor2010, Tumasjan2011,Shi2012,Marchetti-bowick2012,
    Borondo2012,Park2012,Contractor2013,Thapen2013,Hoang2013,Fink2013,Gayo-Avello2013,
    Caldarelli2014,Borge-Holthoefer2015,Tsakalidis2015,Kagan2015,
    Saifuddin2016,Wang2016,Llewellyn2016} also investigated the
  possibility to analyze political processes and predict political
  elections from data collected on Twitter. However, these initial
  investigations achieved only mixed results and engendered a number
  of critical studies \cite{Jungherr2012,Gayo-Avello2013,Jungherr2016}
  questioning their methods and findings. One of the main criticisms
  is that instead of  measuring the political support for a candidate or a party, they
  measure the political attention toward it, those two concepts being
  not necessarily correlated.}
Indeed, most work compare the
volume of tweets, or mentions, related to the different candidates
with traditional polls or election results.
Lexicon-based sentiment
analysis\cite{Subrahmanian2008,Montejo-Raez2014,Tausczik2010} has also
been used to improve this approach by attributing a positive or
negative sentiment to the tweets containing mentions of the candidates
or parties.  However, not only does lexicon-based approach perform poorly
on the informal, unstructured, sometimes ironic, language of
Twitter\cite{Gonzalez-Bailon2015}, but sentiment analysis does not
allow one to differentiate attention from political support, especially
during political scandals\cite{Jungherr2016}.
In this case, correctly capturing the context of the events is crucial to measure supports.

Recent works\cite{Ceron2015,Beauchamp2016} have shown that by going
beyond sentiment analysis, and by considering all the terms used in
tweets, even the terms usually considered neutral, a more accurate
measurement of the Twitter opinion during the 2012 US election was
possible.
Moreover, evidences suggest that it is possible
to differentiate Republican and Democrat Twitter users based only on
their usage of words\cite{Sylwester2015}. 
Ceron {\it et al.}{\high\cite{Ceron2015,Ceron2016,ceron2016politics}} 
used a supervised machine learning approach based on
a hand labeled training set to estimate the proportion of tweets in
favor of each candidate
in the 2012 US election and the 2012 Italian center-left primaries.
Beauchamp\cite{Beauchamp2016} extracted significant textual features
from Twitter by fitting a model to existing polls and showed that
these features improved state level polls prediction.
Despite all these improvements, opinion {\shigh time series} derived 
from Twitter have not been validated so far with any traditional polling performed at
the large scale.

Here, we focus on the 2016 US Presidential Election by collecting
Twitter data regarding the two top candidates to the presidency:
Hillary Clinton (Democratic Party) and Donald J. Trump (Republican
Party).
We develop a supervised learning approach to measure the
opinion of Twitter users where we do not try to classify tweets as
expressing positive or negative sentiment, but as supporting or
opposing one of the candidates.
Our approach innovates by using the network of hashtag
co-occurrence to discover all the hashtags relevant to the elections
and to assess the consistency of our hashtag classification.
This allows us to automatically build a training set approaching one million documents,
which is two order of magnitude larger than what hand labeling typically allows.
Moreover, using an in-domain
training set not only helps us to capture the informalities of Twitter
language, but also permits us to capture the rich context of the 2016
US election.
{\high  We do not attempt to predict the outcome of the elections from
  Twitter, but we show that we can precisely measure the supports of
  each candidate in Twitter, and that while our approach is independent of
  traditional polls, the opinion trend we measure in Twitter closely
  matches the New York Times (NYT) National Polling Average\cite{NYTPolls} and
  anticipates it by several days. The agreement we find significantly
  exceeds results of previous attempts comparing Twitter-based metric
  time series with traditional polls
  \cite{OConnor2010,Marchetti-bowick2012,Thapen2013,Ceron2015,Jungherr2016},
  as we perform a systematic benchmark against all previous methods and show
  that our method outperform all of them.
  By training our model only on the first part of our dataset, we can
  still predict the results of the NYT polls up to 10 days in advance
  during the rest of the election period. We thus validate the use of
  Twitter activity to capture trends existing in the society at the
  national level. We also show that, contrary to the measure of the
  supports of each candidate, the attention toward the candidates, 
  measured by previous studies\cite{OConnor2010,Marchetti-bowick2012,Thapen2013,Jungherr2016},
  does not agree with the NYT national polls.

\label{intro-final}
Finally, by classifying individual users instead of tweets, we
correctly take into account the difference in activity of each users
and we gain unique insight on the dynamics and structure of the social
network of Twitter users in relation to their political opinion. We
show that the difference in behavior and activity between the
supporters of the two candidates results in the fact that Twitter’s
opinion mainly measures the engagement of Clinton supporters. 
This explains the discrepancy between Twitter
opinion and the outcome of the election.
Detecting such a dichotomy before the election is
an important warning signal indicating that the
opinion trend from the polls may not be representative of the electorate.}

\section{Results}

\subsection{Social network of Twitter users}

\begin{figure}[!tb]
\centering
\includegraphics[width=0.9\linewidth]{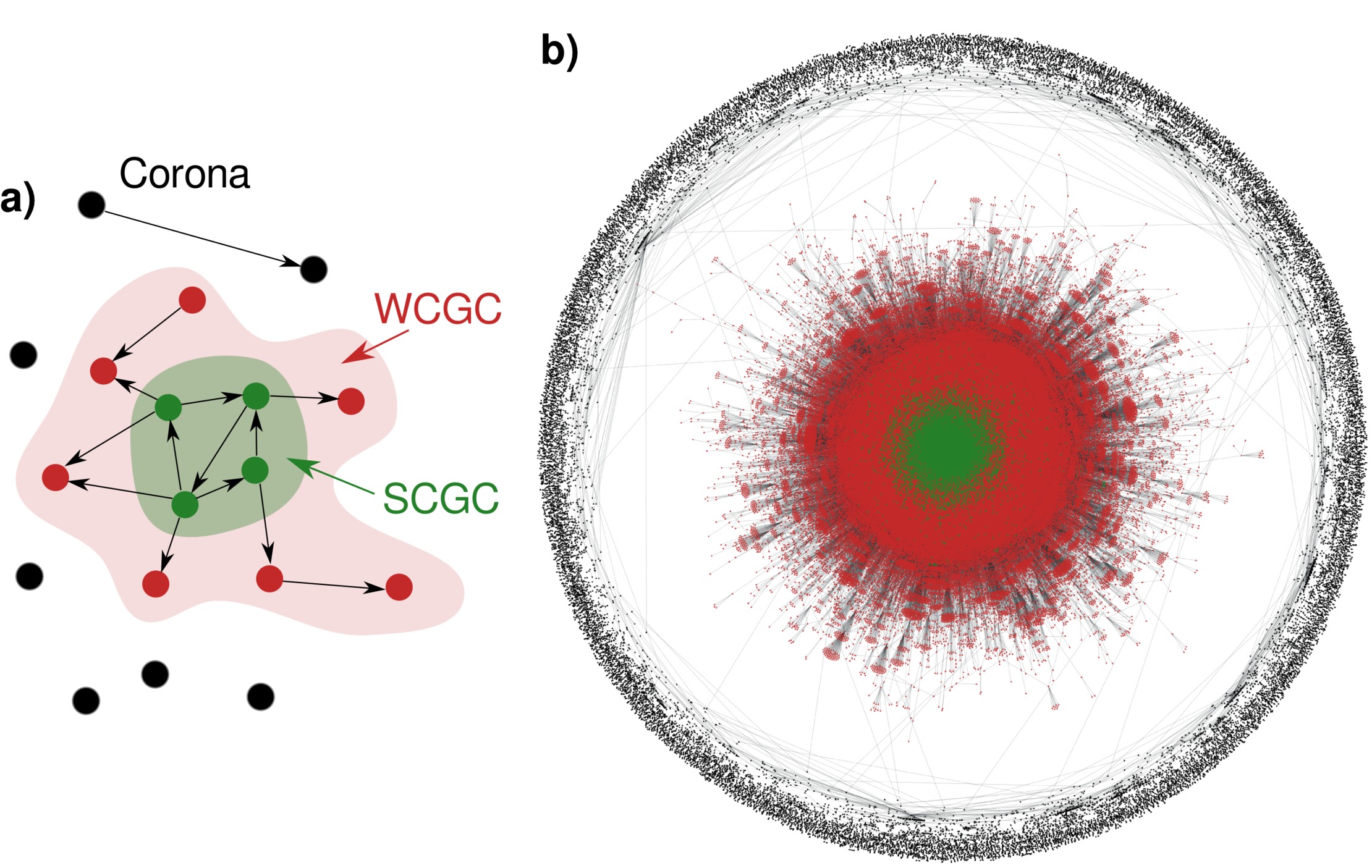}
\caption{{\bf Definition of network components of users of the
    Twitter-election sphere.}  (\textbf{a}) Sketch representing the weakly
  (red) and strongly (green) connected giant components and the corona
  (black).  (\textbf{b}) Visualization of a real influence daily network
  reconstructed from our Twitter dataset.  The strongly connected
  giant component (green) is the largest maximal set of nodes where
  there exists a path in both directions between each pair of nodes.
  The weakly connected giant component (red) is the largest maximal
  set of nodes where there exists a path in at least one direction
  between each pair of nodes.  The corona (black) is formed by the
  smaller components.}
\label{fig:net_comp}
\end{figure}

We collect tweets mentioning the two top candidates in the
2016 US presidential election from June 1st {\shigh until election day on November 8th}, 
2016 by using Twitter Search API to retrieve the following queries:
\textit{trump OR realdonaldtrump OR donaldtrump} and \textit{hillary
  OR clinton OR hillaryclinton}.
The resulting dataset consists of {\shigh 98} million tweets with the keywords about Donald Trump
sent by {\shigh 6.7} million users and {\shigh 78} million tweets with the keywords about Hillary Clinton
sent by {\shigh 8.8} million users.
The combination of the two datasets results in a total of {\shigh 171} million unique tweets.
The total number of users is {\shigh 11 million} with an average of 
{\shigh 1.1 million tweets per day (standard deviation of 0.6 million)
sent by an average of about 375,000 distinct users (standard deviation of 190,000)
per day.}

We then build the daily social networks from user interactions
following the methods developed in Ref. \cite{Pei2014} (see Methods
\ref{sec:net_const}). A directed link between two users is defined
whenever one user retweets, replies to, mentions or quotes another
user.  Using concepts borrowed from percolation
theory\cite{bunde2012fractals,bollobas2001random} we define different
connected components to characterize the connectivity properties of
the network of Twitter users: the strongly connected giant component
(SCGC), weakly connected giant component (WCGC) and the corona (the
rest of the network -- composed of smaller subgraphs not connected
to the giant components SCGC and WCGC).

The SCGC is formed by the users that are part of interaction loops and
are the most involved in discussions while WCGC is formed by users
that do not necessarily have reciprocal interactions with other users
(see Fig. \ref{fig:net_comp}a). A typical daily network is shown in
Fig. \ref{fig:net_comp}b.

\begin{figure}[!tb]
\centering
\includegraphics[width=0.7\linewidth]{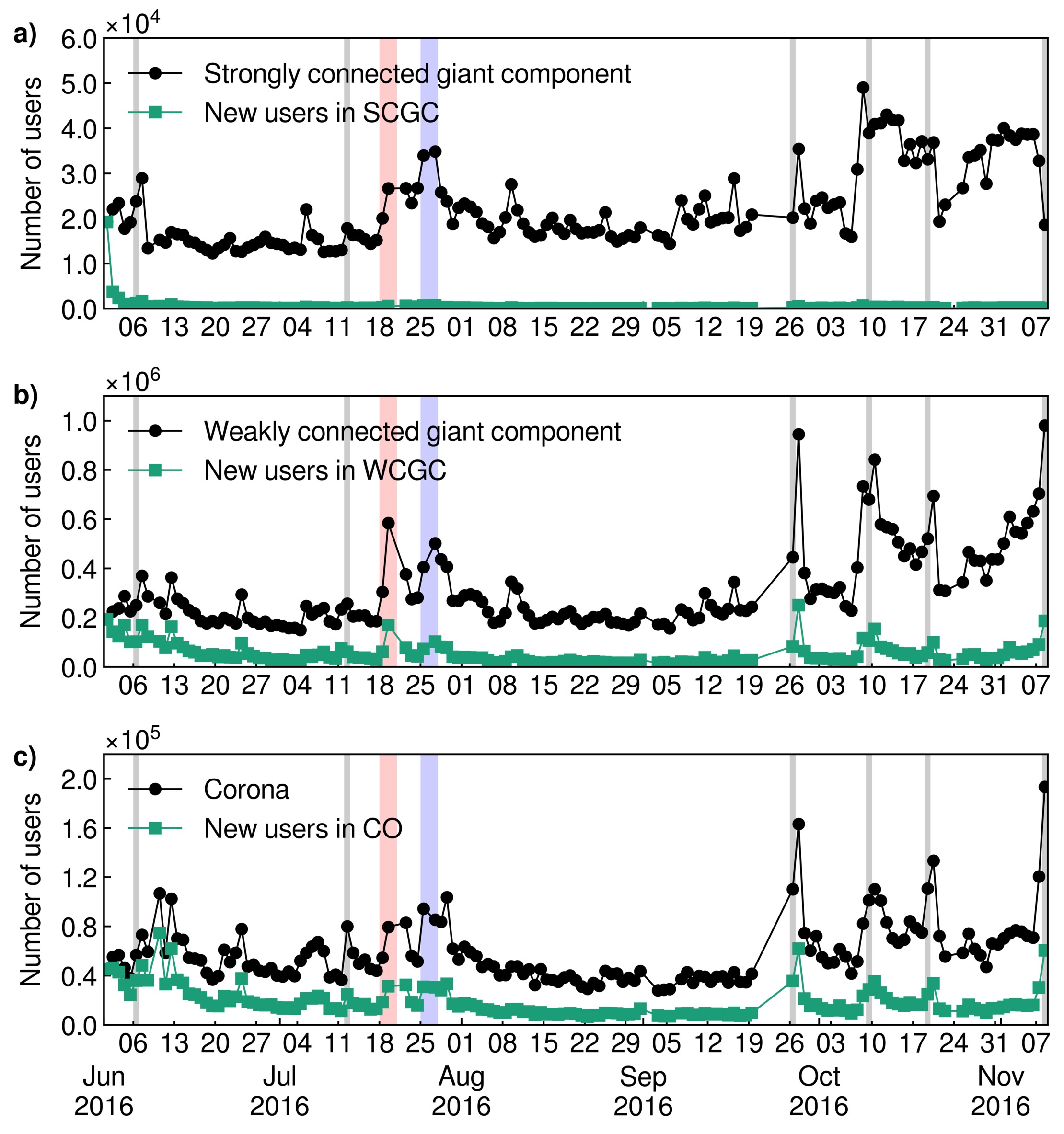}
\caption{{\bf Temporal evolution of daily network components of
    Twitter election users.}  (\textbf{a}) Total number of users in the daily
  strongly connected giant component (SCGC) versus time, (\textbf{b}) weakly connected
  giant component (WCGC) and (\textbf{c}) the corona, i.e. the rest of the components
  (displayed in black in each plot).  The number of new users arriving
  in each compartment is shown in green.  The size of the strongly
  connected component is approximately {\shigh 14} times smaller that the size
  of the weakly connected component.  New users arrive principally in
  the weakly connected giant component or the corona.  The shaded
  areas represents important events: the Associated Press announcing
  of Clinton winning the nomination (June 6), Bernie Sanders
  officially terminating his campaign and endorsing Clinton (July 12),
  the Republican (June 18-21) and Democratic (June 25-28)
  Conventions {\shigh and the three presidential debates (September 26,
  October 9 and October 19).
  Positive fluctuations in the size of the different component
  coincide with these events, in particular for the WCGC and the corona.}}
\label{fig:GC_CO}
\end{figure}

We monitor the evolution of the size of the SCGC, WCGC and the corona
as shown in Fig. \ref{fig:GC_CO}.
{\high 
\label{compfluct}
The WCGC has an average daily size $\simeq$ 310,000 (standard
deviation of 160,000 users) is approximately 14 times larger than the
SCGC with $\simeq$ 22,000 (standard deviation of 8,600) daily users
(see Figs. \FigComponentsEvolution{}a and \FigComponentsEvolution{}b).
The average daily number of users in the
corona is approximately 58,000 with a standard deviation of 25,000
users Fig. \FigComponentsEvolution{}c.}
Fluctuations in the size of the three compartments
are visible in the large spikes in activity occurring during important
events that happened during the period of observation.
For instance,
on June 6, when Hillary Clinton secured enough delegates to be the
nominee of the Democratic Party.
Bernie Sanders (who was the second contender for the Democratic
Nomination) officially terminated his campaign and endorsed Hillary
Clinton on July 12.
The Republican and Democratic Conventions were
held between June 18-21 and June 25-28, respectively
{\shigh and the three presidential debates were held on 
September 26, October 9 and October 19.}
The fluctuations related to these events are more important in the WCGC and
corona than in the SCGC.
The number of new users -- those appearing in our
dataset for the first time -- in each compartment is displayed in green
in Fig. \ref{fig:GC_CO}.  Most new users arrive and connect
directly to the WCGC or populate the disconnected corona while
relatively few users join directly the strongly connected
component.
This is expected as the users belonging to SCGC are those who 
are supposed to be the influencers in the campaigns,
since for users in the SCGC, the information can arrive
from any other member of the giant component, and, vice-versa, the information can flow from the member
to any other user in the SCGC.
Thus, it may take time for a new arrival to belong to the SCGC.
After the first week of observation, the number of new
users arriving directly to the SCGC per day stays stable below
1,000.\\

\subsection{Opinion of Twitter users}

\begin{figure}[!tb]
\centering
\includegraphics[width=\linewidth]{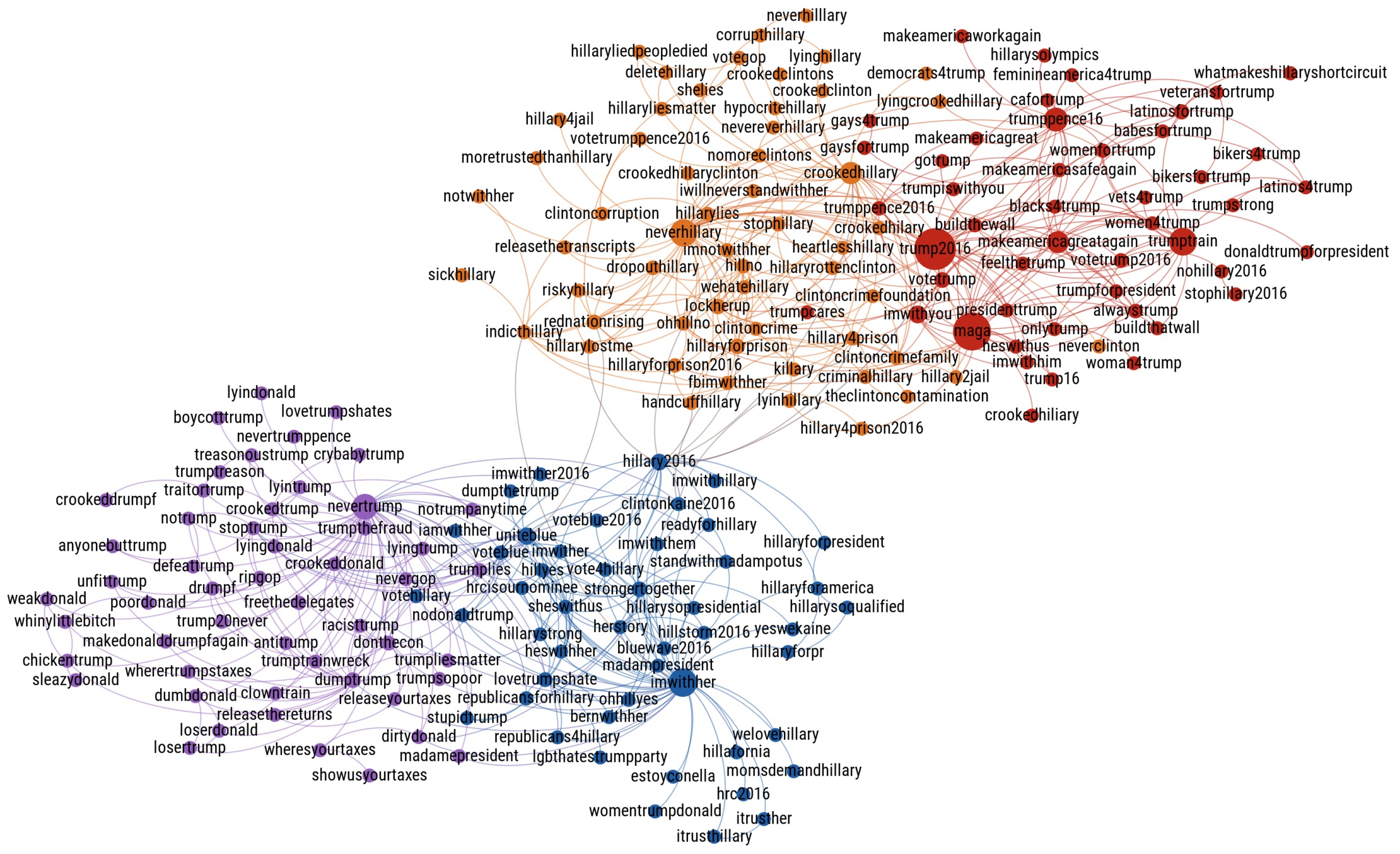}
\caption{{\bf Hashtag classification via network of co-occurrence.}
  {\shigh Network of hashtags obtained by our algorithm from June 1st to September 1st}.
  {\high Nodes of the network represent hashtags and an edge is drawn between two hashtags when
  their co-occurrence in tweets is significant (see Methods \ref{sec:ht_class})}.
  The size of the node is proportional to the total number of occurrence of the hashtag.
  Two  main clusters are visible, corresponding to the
  Pro-Trump/Anti-Clinton and Pro-Clinton/Anti-Trump hashtags.
  Inside of these two clusters, the separation between Pro-Trump (red) and
  Anti-Clinton (orange), or Pro-Clinton (blue) and Anti-Trump
  (purple), is also visible.
  The coloring corresponds to clusters found by community detection\protect{\cite{Raghavan2007,Blondel2008}}.}
\label{fig:ht_cooc}
\end{figure}

We use a set of hashtags expressing opinion to build a set of labeled
tweets, which are used in turn to train a machine learning classifier (see Methods
\ref{sec:ht_class} and \ref{sec:opinion_mining}).
{\shigh Figure
\ref{fig:ht_cooc} displays the
network of hashtags co-occurrence discovered with our algorithm from 
June 1st to September 1st. The network corresponding to the period from
September 1st to November 8th displays similar characteristics and is shown 
in the Supplementary Information.
Hashtags are colored according to the four categories, pro-Trump (red),
anti-Hillary (orange), pro-Clinton (blue) and anti-Trump (purple).
Two main clusters, formed by the pro-Trump and anti-Hillary on the right
and pro-Clinton and anti-Trump on the left, are visible, indicating
a strong relation between the usage of hashtags in these two pairs of
categories.}

The existence of these two main clusters reveals the
strong polarization of the opinion in our dataset and motivates our
decision to reduce our classification of tweets to two categories:
pro-Trump or anti-Hillary and pro-Clinton or anti-Trump, in the
following designated as Trump supporters and Clinton supporters,
respectively.
{\high
\label{ht-sep}
The clear separation of hashtag usage in two main
  clusters corroborates previous studies showing that, in the case of
  political issues, Twitter users exchange primarily among individuals
  with similar ideological preferences as shown in Barbera \textit{et
  al.}\cite{Barbera2015}. This result allows us to use the hashtags from the two
  clusters to create a labeled training set of tweets that is large enough 
  (1 million tweets) so that the opinion of all the Twitter users 
  (11 million) can be inferred with confidence such that it agrees with the NYT 
  National Polling Average.

In principle, it may not be a given fact that any topic of interest
could lead to the same separation in the hashtag network. We could
imagine another topic of discussion in Twitter, let's say, ``Samsung
Galaxy versus IPhone'', where users express their opinion about the
two smartphones. In this case, one would be interested to see whether
the co-occurrence hashtag network separates into two distinct clusters
as in the case of Clinton/Trump. If this is the case, then it implies
that the topic is polarized and our analytics can be applied
confidently to obtain the number of users in favor of one smartphone
or the other. However, it could be the case, that the hashtags do not
separate into two groups. In this case, we would conclude that the
topic is not polarized enough and therefore there are not well defined
groups. Thus, the separation in the hashtag network is a necessary
ingredient that allows to perform the supervised classification.}

Note also that, although Hillary Clinton and Donald
Trump were officially nominated as presidential candidate of their
respective party only during the conventions, they both secured enough
pledged delegates to become the nominee before their party's
convention.  Donald Trump secured enough delegates on May 26 while
Hillary Clinton did it on June 6.

{\shigh Taking into account the entire observation period},
we identify more hashtags in the
pro-Trump {\shigh (n=60)} than in the pro-Hillary {\shigh (n=52)} categories and
approximately the same number in the anti-Trump {\shigh (n=62)} and
anti-Hillary {\shigh (n=65)} categories.  The number of tweets using at least
one of the classified hashtags account for {\shigh 30\%} of all the tweets
containing at least one hashtag.
We find more tweets having hashtags exclusively in the Trump supporters category
than exclusively in the Clinton supporters category ({\shigh 9.6 million} for Trump versus 
{\shigh 3.0} million for Clinton).
These tweets also correspond to more users in the Trump camp than in the Hillary camp
({\shigh 538,720} for Trump versus {\shigh 393,829} for Clinton).
Although these figures might suggest a clear advantage for Trump in Twitter,
one need to take into account 
the whole population of users in the dataset to correctly estimate the popularity of each 
candidate.
This is what we show in section , where the daily opinion of each users is determined 
after having classified our entire dataset of tweets.
\\

{\high
\subsection{Generalization of the method to multi-partite elections and topics beyond elections}
\label{sec:generalization}

A two-classes classification scheme was the best
  approach in the case of the 2016 US Presidential elections since the
  elections where dominated by two candidates. However, in the case of
  multi-partite political systems like some European or Latin American
  countries, we can implement a multi-class classification
  scheme. This is done, for example, generalizing the binary
  classification used with Trump/Clinton to a multi-classification
  scheme involving three or more Parties.

\medskip

The important ingredient for the application of the supervised learning
technique to any kind of topic is the separation of the hashtag
co-occurrence network in well-defined clusters identifying the main
opinions toward the issue at hand. For instance, in the case of
Trump/Clinton, the separation of the co-occurrence hashtag network into
two clear camps allowed the application of the supervised method to the
American election with two main candidates. In the case of the current
French election, we have collected tweets related to
three main candidates of the 2017 French Presidential election
(Fran\c{c}ois Fillon, Marine Le Pen and Jean-Luc M\'elanchon)
to test the generality of our
analytics to this kind of multi-partite political systems.
The chosen candidates correspond to the three main confirmed candidates at the time of data acquisition
(December 19th, 2016 to January 31st, 2017).
This is the reason why we did not acquire data about Emmanuel Macron, who ultimately
passed the first round of the elections along with Marine Le Pen.
The
results of the co-occurrence hashtag networks emanating from the
tweets related to the French elections are shown in Fig. \FigHTNetworkFrench{} in 
the Supplementary Information.
We find that a separation of the
co-occurrence network into three clear clusters is achieved for the
hashtags employed by users expressing supports to three candidates to
the French presidential election. Each group expresses predilection
for each of the three French presidential candidates indicating that
the opinion inference methods can be applied to this kind of
situation as well.

\medskip

Furthermore, we have also investigated whether the method can be
implemented for public opinion outside an election setting. These
``generality tests'' are of importance to distinguish an ad-hoc
research (anecdotal) versus a methodological contribution. For
this purpose, we have collected tweets from a single topic, such as
``climate change'', in search of a generalization of our
algorithms. Figure \FigHTNetworkClimateChange{} in the Supplementary Information shows the result
of the hashtag co-occurrence network in this case. We find that this
network naturally splits into two groups, one with hashtags supporting
action toward climate change and the other with hashtags depicting
climate change as a hoax. This result suggests that our machine
learning and co-occurrence hashtag network method can be generalized
to topics beyond the election setting. The minimal ingredients to
apply our methods are the existence of a set of users interested in
the topic and the appearance of separated hashtag clusters in the
co-occurrence network. This separation was evident in all cases
considered in this study: the US and French elections as well as
climate change.

\medskip

Our approach can thus be extended to understand general trends from
social media including, but not restricted to, societal issues,
opinion on products and brands, to political movements.}

\label{sec:generalization-end}
\subsection{Measuring political support}

\begin{figure}[!tb]
\centering
\includegraphics[width=0.9\linewidth]{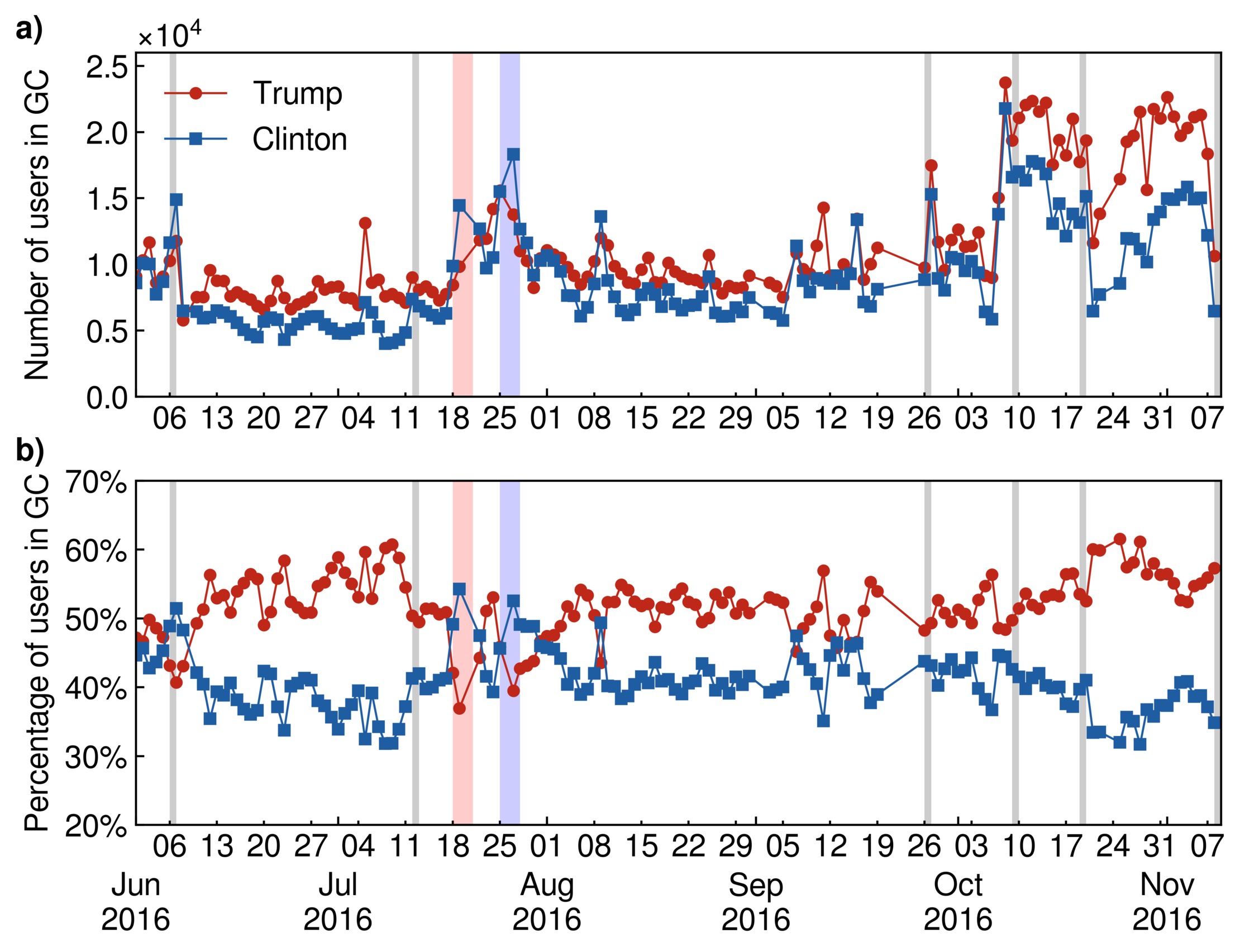}
\caption{{\bf Supporters dynamics in the strongly connected giant component.}
  (\textbf{a}) Absolute number and (\textbf{b}) percentage of supporters of Trump (red,
  Pro-Trump or Anti-Clinton) and Clinton (blue, Pro-Clinton or
  Anti-Trump) inside the strongly connected giant component as a
  function of time.
  {\shigh The opinion in the strongly connected giant
  component is clearly in favor of Donald Trump and
  shifts slightly in favor of Hillary Clinton only
  occasionally, such as during the two Conventions.}
  In (\textbf{b}) the data adds to 100\% when considering the
  unclassified users ($\simeq$ 4.5\%).
  {\shigh Donald Trump's advantage in the SCGC
  reverses when including the opinion of all Twitter users shown in
  Fig. \ref{fig:num_users}b.}}
\label{fig:num_users_in_gc}
\end{figure}

\begin{figure}[!tb]
\centering
\includegraphics[width=0.9\linewidth]{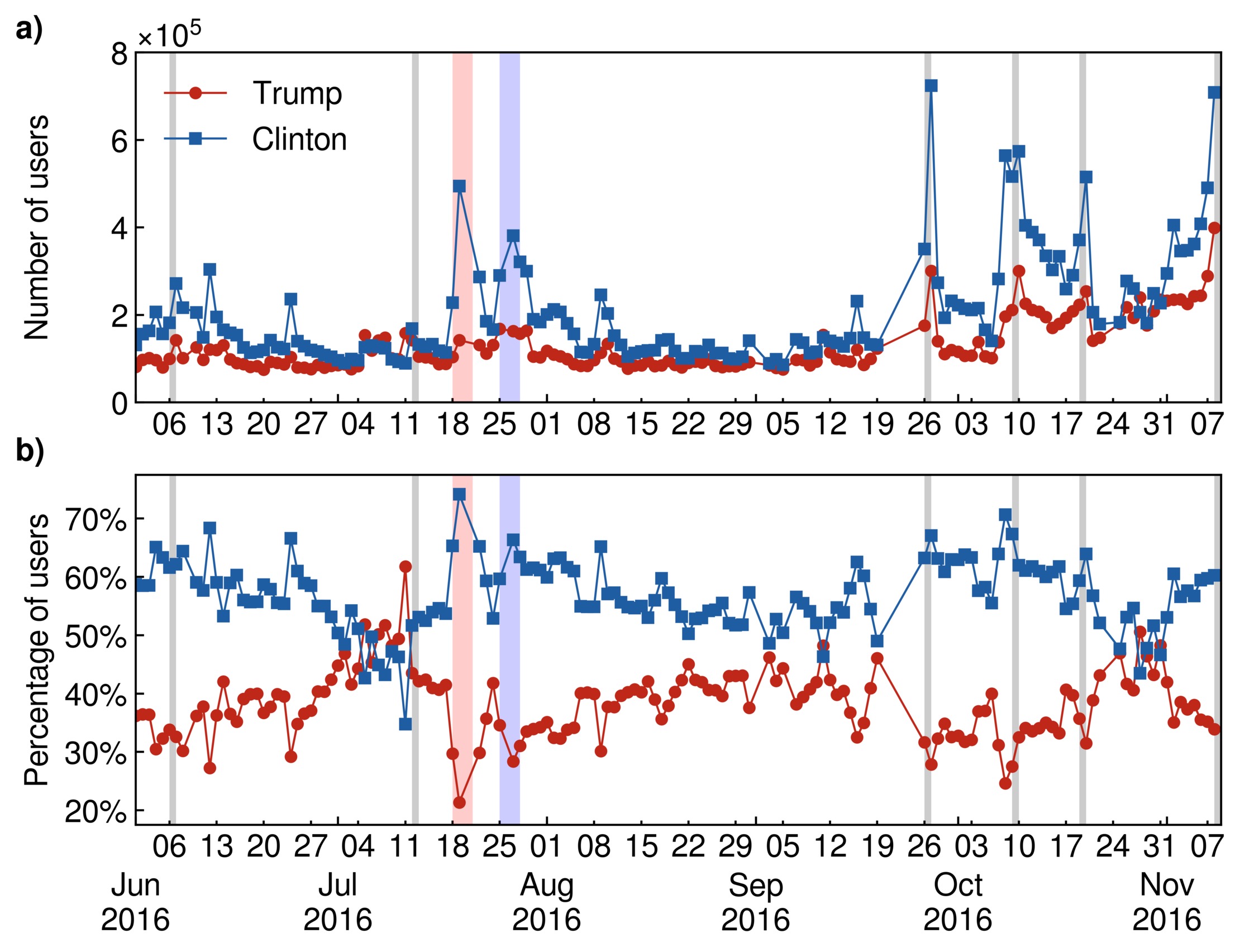}
\caption{{\bf Supporters in the entire Twitter dataset.}  (\textbf{a})
  Total number and (\textbf{b}) percentage of users labeled as Trump (red) and
  as Clinton (blue) in our entire Twitter population as a function of time. 
  Taking into account all
  the users in our dataset, the popular opinion is generally strongly in favor
  of Hillary Clinton in contrast with the strongly connected giant
  component in Fig. \ref{fig:num_users_in_gc}b.
  {\shigh The popularity of
  Donald Trump peaks before the conventions and before the election, however
  Hillary Clinton dominates Twitter opinion, in particular 
  during important events, such as the conventions, the presidential
  debates and the election, coinciding with large positive fluctuations
  in the total number of users.}}
\label{fig:num_users}
\end{figure}

The absolute number of users expressing support for Clinton and Trump
as well as relative percentage of supporters for each party's candidate
in the strongly connected component and in the entire population
dataset are shown in Figs. \ref{fig:num_users_in_gc} and
\ref{fig:num_users}, respectively.  Results for the weakly connected
component are similar to the whole population.  The support of each
users is assigned to the candidate for which the majority of its daily
tweets are classified (see Methods \ref{sec:opinion_mining}).
Approximately 4.5\% of the users are unclassified every day, as they
posts the same number of tweets supporting Trump and Clinton.
We only consider tweets
originating from official Twitter clients in order to discard tweets
that might originate from bots and to limit the number of tweets
posted from professional accounts.  After this filtering, 92\% of the)
total number of tweets remain.

We find important differences in the popularity of the candidates
according to the giant components considered.  The majority of users
in the SCGC are clearly in favor of Donald Trump for the majority 
of the time of observation (Fig. \ref{fig:num_users_in_gc}).
However, the
situation is reversed, with Clinton being more popular than Trump,
when the entire Twitter dataset population is taken into account
(Fig. \ref{fig:num_users}), revealing a difference in the network
localization of the users belonging to the different political
parties.
{\shigh A difference in the dynamics of the supporters' opinion is
also uncovered: during important events, such as
the conventions or the presidential debates, Hillary Clinton's supporters show
a much more important response than Donald Trump's supporters (Fig. \ref{fig:num_users}a)
and even sometimes slightly dominate the SCGC (Fig. \ref{fig:num_users_in_gc}b).
This difference in behavior is also manifested in the fact
that spikes in favor of Donald Trump in the percentage of opinion (Fig. \ref{fig:num_users}b),
such as on October 28 when FBI Director, James B. Comey, sent a letter to the Congress saying that new emails,
potentially linked to the closed investigation into whether Hillary Clinton had mishandled classified 
information, had be found,
correspond rather to a lack of activity of Clinton's supporters then to an increase in the
engagement of Trump's supporters (seen in Fig. \ref{fig:num_users}a).}
{\shigh We analyze these differences in behavior and their impact on the Twitter
opinion trend in section \ref{sec:twitter_behav}}.\\

{\high
\label{scgcvswhole}
The inversion of the opinion of the SCGC as compared
  with the result of the whole network 
  allows us to understand the behavior of Trump/Clinton voters in a way
  that only Twitter can, since the network information is only
  available from Twitter but not from the National Polls. That is, the
  Trump supporters are the majority in the SCGC, while the Clinton
  supporters are the majority in the whole network. This last result
  is the only one that agree with the NYT National Polls. Using the
  entire dataset allows us to capture the opinion of the full
  Twitter population, which is what we then find in agreement with the
  National Polls.}

\subsection{Comparison with national polls aggregates}
\label{sec:fit_nyt}
We next compare the daily global opinion measured in
our entire Twitter dataset with the opinion obtained from traditional
polls.  We use the National Polling Average computed by the New York
Times (NYT)\cite{NYTPolls}
which is a weighted average of all polls (n=270) listed in the
Huffington Post Pollster API
(\url{http://elections.huffingtonpost.com/pollster/api}).  Greater
weights are given to polls conducted more recently and polls with a
larger sample size.  Three types of traditional polls are used: live
telephone polls, online polls and interactive voice response polls.
The sample size of each poll typically varies between several
hundreds and tens of thousands respondents and therefore the aggregate
of all polls considered by the NYT represents a sampling size in the
hundred of thousand of respondents.

\begin{figure}[!tb]
\centering
\includegraphics[width=0.9\linewidth]{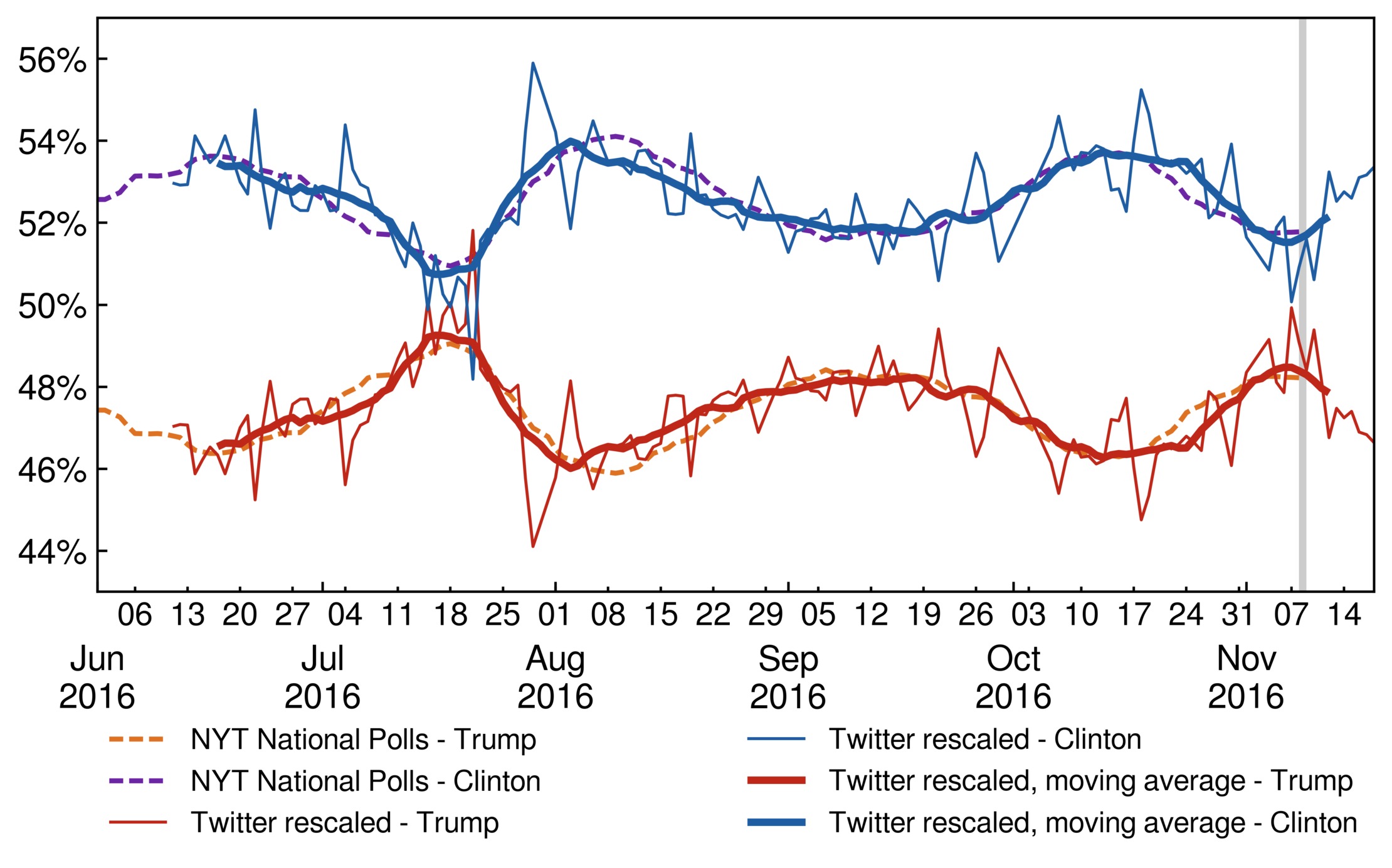}
\caption{{\bf Validation of Twitter election trend and NYT aggregate
    national polls.} {\high Least square fit of the percentage of Twitter
  supporters in favor of Donald Trump and Hillary Clinton with the
  results of the polls aggregated by the New York Times for the
  popular votes and normalized to the share of the two candidates.
  Twitter opinion time series are in close agreement
  with the NYT National Polls.
  As Twitter provides an instantaneous
  measure of the opinion of its users, a time-shift of 10 days exists between the
  New York Times polls and the Twitter opinion.
  Pearson's coefficient between the NYT and the 13 days
  moving averaged Twitter opinion has a remarkably high value $r\simeq0.93$
  with a root-mean-square error (RMSE) of $\simeq 0.31$\,\%.
  Values of the correlation coefficient and RMSE as a function of the window
  averaging size are displayed in Fig. \FigNYTFitRMSER{}}}
\label{fig:nyt_fit}
\end{figure}

\begin{figure}[!tb]
\centering
\includegraphics[width=0.5\linewidth]{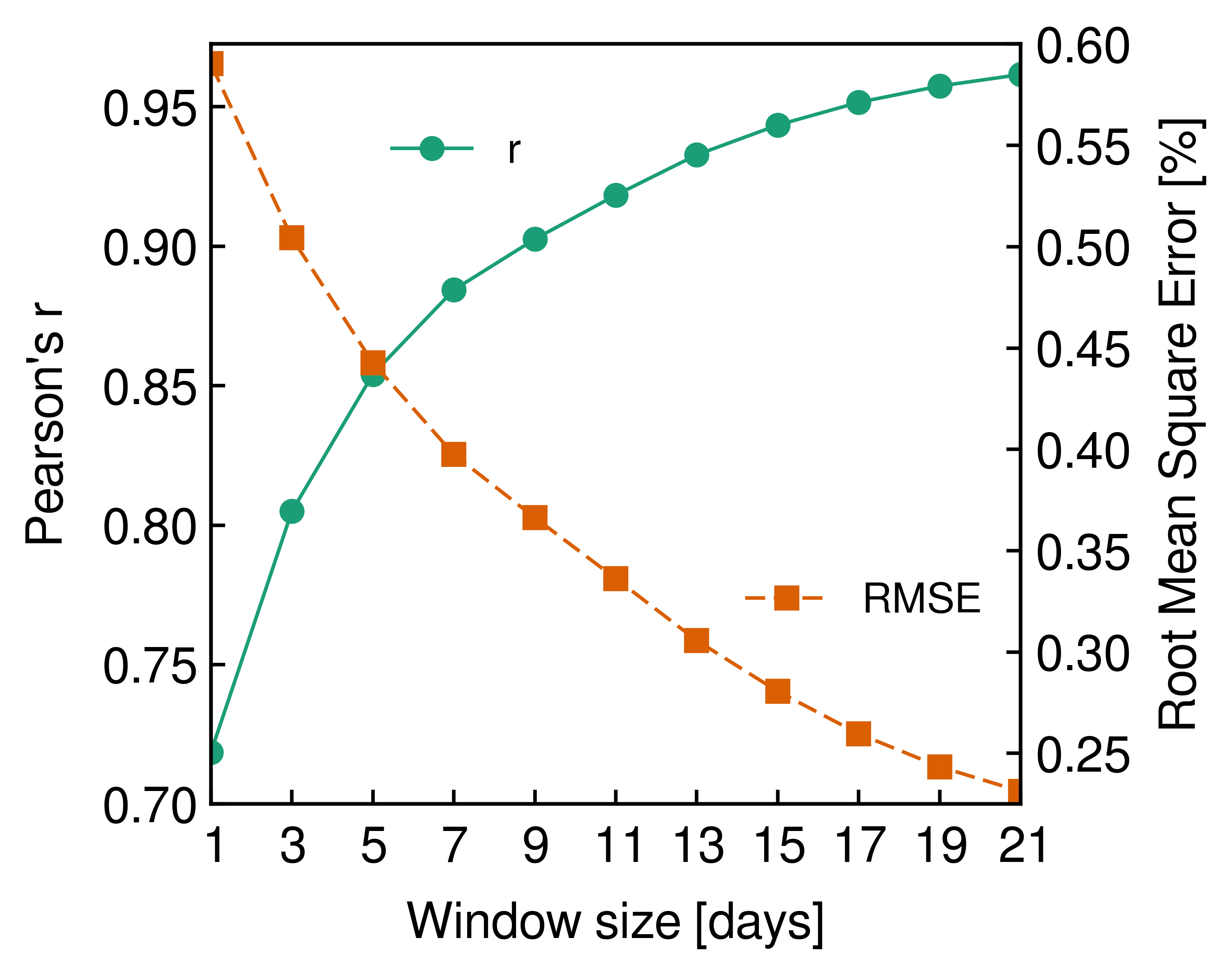}
\caption{\high {\bf Agreement of the fit for different window averaging size.}
  Pearson correlation coefficient ($r$) and root-mean-square error (RMSE) of the fit as 
a function of the moving average window length.
The Pearson coefficient, quickly increases and the root-mean-square error 
decreases as the window 
length increases and smooths out the daily fluctuations.
The best fit is obtained for a window length of 21 days with RMSE\,$\simeq0.23$\% and $r\simeq0.96$.
RMSE is expressed in percentage point of the NYT polls.}
\label{fig:fit_rmse_r}
\end{figure}

{\high
\label{normalization}
We remove the shares of undecided and third party
  candidates from the NYT polling average and compare the resulting
  relative opinion trend with the ratio of Twitter users in favor of
  Donald Trump or Hillary Clinton.}
The comparison between our Twitter {\shigh opinion}, Fig.
\ref{fig:num_users}b, and the New York Times national polling average
is shown in Fig. \ref{fig:nyt_fit}.  The global opinion obtained from
our Twitter dataset is in excellent agreement with the NYT polling
average..

The scale of the oscillations visible in the support trends in Twitter
and in the NYT polls are also in agreement beyond the small scale
fluctuations which are visible in the Twitter opinion time series
since it represents a largely fluctuating daily average.  Furthermore,
a time shift exists between the opinion in Twitter and the NYT polls
in the sense that the Twitter data anticipates the NYT National Polls
by several days. This shift reflects the fact that Twitter represents
the fresh, instantaneous opinion of its users while traditional polls
may represent a delayed response of the general population that takes
more time to spread, as well as typical delays in performing and
compiling traditional polls by pollsters.

{\shigh In order to precisely evaluate the agreement between Twitter and
NYT time series, we perform a least square fit of a linear function
of the Twitter normalized ratio of supporters of
each candidate to their NYT normalized popularity percentage
(see Methods \ref{sec:fit_methods}).
Specifically, we apply the following transformation:

\begin{equation}
{r'}_w^k(i) \mapsto A^k \,\, r_w^k(i-t_d) \, +\, b^k,
\label{eq:fit}
\end{equation}

where $r_w^k(i)$ is the ratio of Twitter users in favor of a candidate $k$
at day $i$ to which we applied a backward moving average with 
a window length $w$.
The rescaling parameters $A^k$ and $b^k$ are 
the parameters that fit the NYT polls and $t_d$ is a time delay between the Twitter 
opinion and the polls.
The moving window average of $w$ days
converts fluctuating daily data into a smooth trend that can be
compared with the NYT smooth time series aggregated over many
polls performed over several days.
{\high 
\label{nofuture}
We use a backward window average to ensure
that no data from the future is used.}
Note that a backward moving average induces an
artificial backward time shift of $(w-1)/2$
(see Methods \ref{sec:fit_methods}) so that 
the full time shift between the Twitter time-series
and the NYT polls that we report below is given by $T_d = t_d + (w-1)/2$.
}

{\shigh Figure \ref{fig:nyt_fit} shows the fit using a window averaging 
of the Twitter data of 13 days.
The constant parameters that provide the best fit in this case 
are $A^C=A^T=0.185$, $b^C= 1 - b^{T} - A^{T}=0.415$, $T_d = 10$\, days.
The remarkable agreement of the fit is characterized by a
Pearson product-moment correlation coefficient $r = 0.93$
and a root-mean-square error of RMSE = 0.31\,\%, expressed in
percentage points of the NYT polls.
Using longer window average length increases the quality of the fit
as shown in Fig. \ref{fig:fit_rmse_r} displaying the root-mean-square error expressed
in percentage points of the NYT polls and the Pearson correlation
coefficient of the fit as a function of the moving average window length $w$.}

It is important to note that Twitter data cannot predict the exact percentage of supporters to each
candidate in the general population due to the uncertainty about the
number of voters that do not express their opinion on Twitter and
about the number of users that are undecided and are not classified by
our machine learning. 
However, it is more important to capture the
trend of both candidates' popularity in respect to each other,
which is obtained from Twitter.  Even if Twitter may not provide the
exact percentage of support for each candidate nationwide, the
relevant relative opinion trend is fully captured by Twitter with
precision. Furthermore, the important parameter is $T_d$, the time
delay between the anticipated opinion trend in Twitter and the delayed
response captured by the NYT population at large.
{\shigh We find that this delay time is independent from the actual 
value of the popularity of each candidate and from the length of the 
window average.}\\

{\high
\label{prediction}
Next, we investigate how well our Twitter opinion
  trend can predict the NYT National Polling Average. Using the best
  fitting parameters to predict the NYT Polls from our Twitter opinion
  in a post hoc manner is a flawed approach since it requires knowing
  the entire time series. To remedy this issue and correctly test the
  predictive power of our Twitter analysis, we train our entire model,
  i.e. we train our classifier using tweets labeled with hashtags
  found with the hashtag co-occurrence network, compute the daily
  ratio of users in favor of each candidate and find the parameters
  $A$, $b$ and $T_d$ that best fit the NYT polls, using only the
  portion of our data ranging from June 1st until September 1st. We
  then classify with the rest of the tweets the pre-trained classifier
  and keep the same fitting parameters to compare our Twitter trend
  with the NYT polls until election day on November 8th. We take
  advantage of the fact that the Twitter opinion time series precedes
  the NYT polls by several days to predict future values of the
  NYT polls using only data from the past. We compare this results
  with a straightforward extrapolation in time of the NYT polls
  from a linear regression on the last three weeks of the polls, as in
  Beauchamp\cite{Beauchamp2016}, and constant extrapolation using the mean value of the polls
during the training period.

Training our model only on the first three months,
the time delay giving the best fit is $T_d=11$\,days.
We use our Twitter trend smoothed with a 9 days
backward moving average to predict the polls 
7 days into the future (see Methods).
We find that our Twitter opinion predicts the NYT polls with better
accuracy, i.e. a smaller root-mean-square error (RMSE), than a
straightforward linear extrapolation and 
a simple constant extrapolation using the mean value of the polls
(see Fig \FigPollPrediction{}).
Predicting the polls 7 days in
advance, we find that the Twitter prediction reduces the RMSE
by 52\% compared to the constant extrapolation 
and by 66\% compared to the linear extrapolation.
The prediction error of the different methods are displayed
in Fig. \FigPollPrediction{}b.
Even more importantly, our Twitter opinion is particularly
better than the linear extrapolation to predict rapid changes
in the polls as seen in Fig. \FigPollPrediction{}a which shows the prediction
from Twitter along with the result of the linear extrapolation.
Although the linear extrapolation unsurprisingly predict relatively well
the polls when they undergo small variations, 
when the polls experience a trend reversal, 
the Twitter prediction accurately predict it while the linear 
extrapolation misses it.
The ability of our method
to predict the polls during more than two months without 
reusing the polls to retrain our model, serves as a additional validation of the opinion
we measure in Twitter.
}

\begin{figure}[!tb]
\centering
\includegraphics[width=0.9\linewidth]{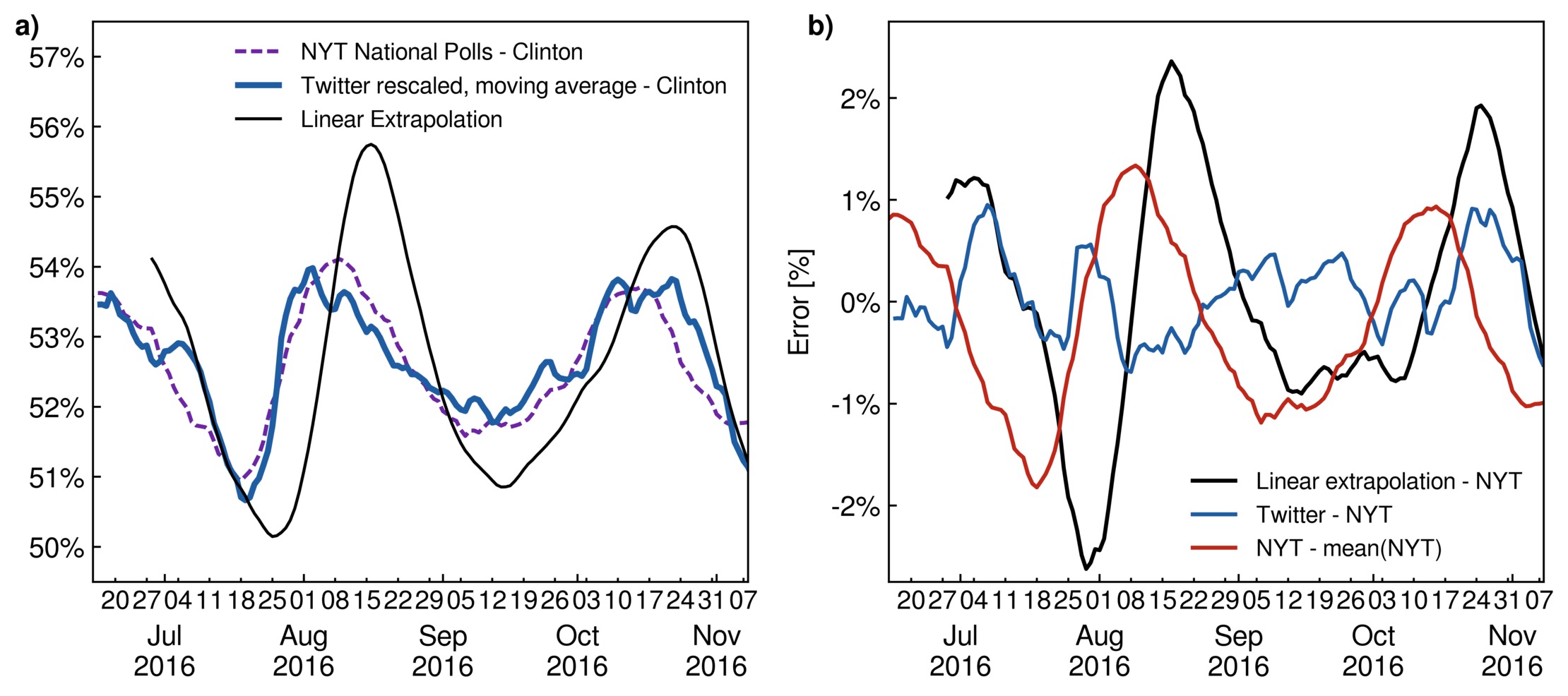}
\caption{{\high {\bf Twitter 7 days prediction versus linear extrapolation of the polls.}
(\textbf{a}) Twitter prediction of the NYT polls 7 days in advance (blue line),
7 days linear extrapolation of the NYT polls (black line) and 
Hillary Clinton NYT National Polling Average score,
normalized to the share of Donald Trump and Hillary Clinton,
(dashed purple line).
Our model is trained using only data from June 1st to September 1st.
(\textbf{b}) Prediction error in percentage points of the NYT polls.
The Twitter prediction error (blue) has a root-mean-square value of RMSE = 0.40\%
(correlation coefficient $r = 0.89$).
The 7 days linear extrapolation of the polls (black) has a RMSE = 1.19\% ($r = 0.64$) 
and the baseline error, computed as the difference between the NYT Polling Average and the its mean value (red),
achieves a RMSE = 0.83\%.
}}
\label{fig:nyt_partial_fit}
\end{figure}

\label{prediction-end}

{\high \subsection{Analysis of Twitter supporter behavior}
\label{sec:twitter_behav}
 We showed that the variations of opinion measured in
  Twitter are in very good agreement with the variations of relative
  opinion in the NYT National Polling average, and, as for the NYT
  polls, the majority in Twitter is in favor of Hillary Clinton.
  However, traditional polls generally failed at predicting the
  outcome of the 2016 presidential election. 
An interesting question that arises is then:  
  is there any warning
  signals in the Twitter data that allows us to predict in
  advance that there could be a surprising result at the election day?

Here, we show that
analyzing the behavior of supporters in Twitter can be used to detect
such problems. This analysis is not accessible to traditional polling
and, given the agreement between Twitter and traditional polls, might
serve as a warning for a possible similar problem in
traditional polls and their subsequent failure to predict the elections.\\

Firstly, in addition to measuring the daily opinion, we measure the
opinion of the entire population of Twitter users whose tweets we
collected over the period going from June 1st until November 8th,
something not possible for the majority of traditional polls. That
is, using all the tweets in our dataset posted by a users over the
entire observation period, we classify each users according to camp in
which the majority of his/her tweets is classified. This
calculation contrast to the one employed by polls, which can track
only a sample population at a given time. In this cumulative count,
each user is only counted once, while in the daily count, a user is
counted every day she/he expresses her/his opinion in Twitter. Considering
this cumulative count, we find that a large majority of users, 64\%,
is in favor of Hillary Clinton while 28\% are in favor of Donald Trump
and 8\% are unclassified as they have the same number of tweets in
each camp.
The prediction of the cumulative count contrasts with the prediction of the 
daily count.
The average of the daily opinion over the same period
(Fig. \FigDailyOpinion{}) amount to 55\% for Hillary Clinton versus 40\%
for Donald Trump (5\% unclassified). Such a large difference between
the daily and the cumulative ratios of supporters is our first warning
signal indicating a problem in the representation of both supporters
in Twitter. Indeed, looking at the activity of users in both camps we
see that Trump supporters are, in average, much more active. Figure
\FigUsersActivity{}a shows the daily average of tweets per users in
each camp. Clinton supporters tweet in average 2.6 times per day
while Trump supporters tweet in average 3.9 times per day and their
activity increases to almost 6 tweets per day during the period of the
presidential debates.\\

The cumulative distribution of the number of times a user tweets for
each camp (Fig. \FigUsersActivity{}a) reveal a clear difference in the
activity profiles of supporters of each camps. While both
distributions follow a power-law form with a soft cut-off starting
around 1000 tweets per users, the distribution of Trump users shows a
less steeper slope than the one for Clinton supporters revealing
that Trump supporters are generally characterized by a larger activity
and that a small number of Trump supporters have a extremely high
activity. Figure \FigUsersActivity{}b shows that by considering only
users for which we collected at least 67 tweets during the entire
observation period, the advantage tilts in favor of Donald Trump.
\label{comp-discrepancy}
This discrepancy in activity between supporters is also apparent
in the structure of the social network where the strongly connected
giant component, which show less size fluctuations than the other
components and is mainly comprised of recurring users, is dominated by
Trump supporters (Fig. \FigDailyOpinionSCGC{}) in clear opposition to the entire dataset
(Fig. \FigDailyOpinion{}).\\

A second observation available from our Twitter analysis that is not
available to traditional pollsters is the difference in the dynamics
of supporters of each camp. Figure \FigDailyOpinion{}a shows that the
daily number of Trump supporters fluctuates less than the number of
Clinton supporters. We find that $\sigma_{n_C}/\sigma_{n_T} \simeq
2.1$ where $\sigma_{n_k}$ is the standard deviation of the daily
number of users in favor of user $k$. Trump supporters show a more
constant supports while Clinton supporters show their supports mainly
when important events occurs, leading to larger fluctuations in their
daily absolute number. To understand the impact of these different
behaviors on the value of the ratio of users in favor of each
candidate, we evaluate Spearman's rank correlation coefficients
between the daily value of the absolute number of users in favor of a
candidate (Fig. \FigDailyOpinion{}a) and the ratio of user in favor of
the same candidate (Fig. \FigDailyOpinion{}b). We find a value of
$\rho_C \simeq 0.72$ for Clinton supporters and $\rho_T \simeq -0.28$.
This results in the important fact that the relative variations of the
daily opinion that we measure on Twitter (Fig. \FigDailyOpinion{}b),
which agrees with the NYT polling average, are mainly explained by the
variation of the support of Clinton supporters and almost not by the
variation of the support of Trump users. Moreover, the negative value
of $\rho_T$ indicates that a positive fluctuation in the number of
Trump supporters is generally correlated with even larger increase
in the number of Clinton supporters. This analysis shows how opinion
trends measured in Twitter can be understood as the results of the
dynamics of the different supporters camp. The opinion trend mainly
reflect the daily fluctuations of the Hillary supporters coming in and
out of the sampled population and misses the response of Trump
supporters. As our Twitter opinion correlates very well with the
traditional polls, it can be interpreted as a warning for similar
demographic problems in the traditional polls.
}

\begin{figure}[!tb]
\centering
\includegraphics[width=0.9\linewidth]{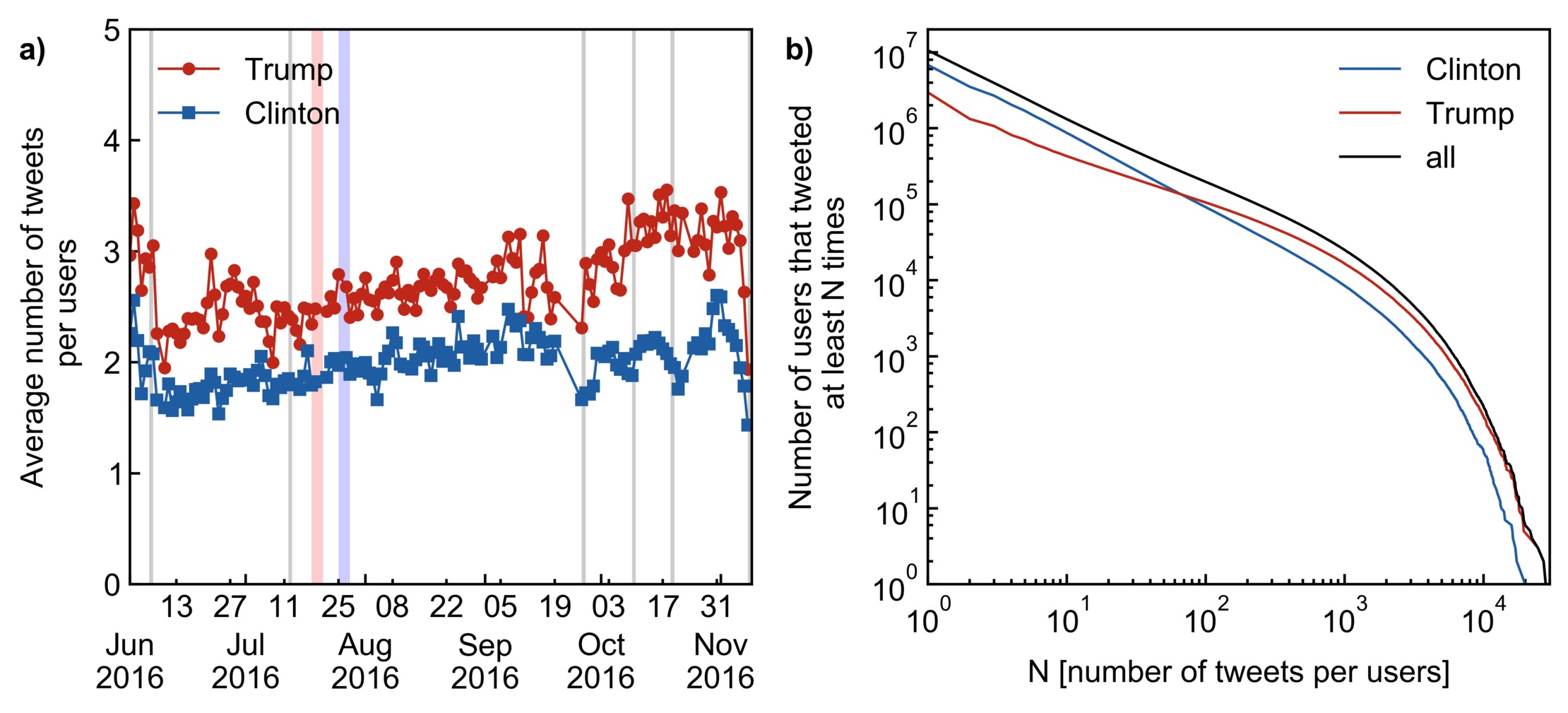}
\caption{\high {\bf Activity of Twitter supporters.}
{\bf (a)}  Daily average number of tweets per user of each camp.
Trump supporters have a higher average activity (shown in red), tweeting on
average 3.9 times per day while Clinton supporters tweet
on average 2.6 times (show in blue).
{\bf (b)} Distribution of user activity.
Activity of all users (black), Clinton supporters (blue) 
and Trump supporters (red).
The distributions are characterized by a power-law shape with a soft
cut-off around 1000 tweets per users. 
The distribution for Clinton supporters has a steeper power-law
indicating a generally smaller activity than Trump supporters.
}
\label{fig:num_tweets_per_user_CCDF}
\end{figure}

\label{sec:twitter_behav-end}

{\high
\subsection{Benchmark with other Twitter-based metrics}
\label{sec:benchmark}

Here we compare the performance of our method
with the approaches used previously.
Similarly to the approach we used
to fit our Twitter opinion time series
to the normalized NYT national 
polling average scores of the two main candidates,
we build metrics $M^k_l$ for each approach $l$, where
$k\in\left(C,T\right)$ represent the candidate ($C$ for Clinton
and $T$ for Trump), 
such that $M^C_l(i) + M^T_l(i) = 1$ for each 
day $i$.
Since the metrics are 
complementary in respect to each candidate,
we only need to compare the metric for one candidate
with the poll scores of the same candidate.

The first approach consists in simply counting the number of users mentioning
each candidate per day. 
This approach used by many authors(e.g. \cite{OConnor2010,Tumasjan2011,
Fink2013,Caldarelli2014,Jungherr2016,Saifuddin2016}) is generally
thought to measure attention toward a candidate rather than 
opinion\cite{Gayo-Avello2013,Jungherr2016}.
O'Connor \textit{et al.}\cite{OConnor2010} reported a correlation 
of $r = 0.79$ between
the number of tweets per day (using a 15 days window average) 
mentioning Barack Obama and his score in the polls
during the 2008 US Presidential elections.
However, the authors found that the McCain 15-day mention volume
also correlated to higher Obama ratings.
Jungherr \textit{et al.}\cite{Jungherr2016} reported 
correlations between the number of mentions per day
of different parties and their polls scores during the 2013 German federal elections.
The largest correlation being $r = 0.279$ for the party
``Alternative f\"{u}r Deutschland'' (with a time lag of 1 day between the polls and the Twitter metric).
We compare the time series given by
\begin{equation}
 M^C_{\textrm{mentions}}(i) = \frac{N^C_u(i)}{N^C_u(i)+N^T_u(i)},
 \label{eq:mentions}
\end{equation}
where $N^k_u(i)$ is the number of users mentioning candidate $k$ during day $i$,
with the normalized poll score of Hillary Clinton.
We use the keywords \textit{donald}, \textit{trump}, \textit{donaldtrump} and \textit{realdonaldtrump}
to count mentions of Donald Trump and \textit{hillary}, \textit{clinton} and \textit{hillaryclinton}
for mentions of Hillary Clinton.

The second metrics we use consists of adding a sentiment analysis to the mention counts.
This approach has also been wildly used (e.g. \cite{OConnor2010,Shi2012,Marchetti-bowick2012,
Thapen2013,Ceron2015,Jungherr2016,Saifuddin2016}) by inferring the sentiment of a tweets
using lexicons\cite{OConnor2010,Thapen2013,Saifuddin2016} or supervised-learning\cite{Marchetti-bowick2012,Ceron2015}.
O'Connor \textit{et al.}\cite{OConnor2010} reported a smaller correlation for Obama
taking into account sentiment ($r = 0.44$) compared to just counting mentions
and a correlation of $r=0.731$ for the sentiment of the keyword \textit{jobs}
with the time series of the consumer confidence (using a 15 days window average).
For comparing with the polls, we define the metrics
\begin{equation}
 M^C_{\textrm{ mentions-emotion}}(i) = \frac{N^\textrm{C,pos}_u(i)+N^\textrm{T,neg}_u(i)}{N^\textrm{C,pos}_u(i)+N^\textrm{C,neg}_u(i)+N^\textrm{T,neg}_u(i)+N^\textrm{T,pos}_u(i)},
\label{eq:mentions-sentiment}
 \end{equation}
where $N^{k,e}_u(i)$ is the number of users that mentioned candidate $k$ in a tweet with sentiment $e$.
To infer the sentiment $e\in\left(pos,neg\right)$ of a tweet we trained a classifier on a training set
comprising tweets from our datasets with positive and negative emoticons and emojis.
This is similar to the method used in Ref. \cite{Marchetti-bowick2012}.
We use supervised learning instead of a lexicon based approach due to the poor 
performance of such approach on the informal text of tweets\cite{Gonzalez-Bailon2015}.
We use a tweet-level classification instead of an approach
allowing to infer directly the aggregated tweet sentiment values\cite{Hopkins2010,Ceron2015,Ceron2016}
in order to be able to compute the user ratio in each camp, required to compare with our results.

The third metrics we consider is derived from the number of hashtags
referring to the candidates.
We define the metric as
\begin{equation}
 M^C_{\textrm{hashtags}}(i) = \frac{N^\textrm{pro-C}_u(i)+N^\textrm{anti-T}_u(i)}{N^\textrm{pro-C}_u(i)+N^\textrm{anti-C}_u(i)+N^\textrm{pro-T}_u(i)+N^\textrm{anti-T}_u(i)},
\label{eq:hashtags}
 \end{equation}
where $N^\textrm{pro-k}_u(i)$, respectively $N^\textrm{anti-k}_u(i)$, is the number 
of users using at least one hashtag in favor of, respectively in opposition to,
candidate $k$ during day $i$.
To represent each category, we use hashtags chosen among the top used hashtags:
\textit{\#MAGA} for pro-Trump, \textit{\#ImWithHer} for pro-Clinton, \textit{\#NeverTrump} for anti-Trump
and \textit{\#NeverHillary} for anti-Clinton.
This set is the same set that we used as a seed 
in the co-occurrence hashtag network for our hashtag discovery algorithm
(see Methods).
By counting hashtags and hashtags with positive or negative values
a party, Jungherr \textit{et al.}\cite{Jungherr2016} reported 
a maximum absolute correlation of $r=-0.564$ for the party ``Die Gr\"{u}ne''
during the German 2013 federal elections (with a negative 
lag of 1 day between the polls and the Twitter time series).

Figure \ref{fig:fit_polls_wl13_benchmark} shows
the results of the three time series
obtained with these metrics,
along with the result of our Twitter opinion 
time series, using a moving window average of 13 days in each case.

\label{bench-discus}
 In all cases, the quality of the agreement that we
  find using our analytics, as expressed in
  Fig.~\FigCompBenchmark{}, between the national polls
  and Twitter trends is superior to previous approaches.

The other methods benchmarked agree poorly with the 
National Polls, and, as can be seen in Fig. \FigCompBenchmark{}.
This evidence shows that the opinion in Twitter is different than the attention,
which is what was measured in previous
studies\cite{OConnor2010,Marchetti-bowick2012,Thapen2013,Ceron2015,Jungherr2016}.
These results confirm the importance of correctly measuring opinion in
Twitter by assessing the supports of each user, something we achieve
using a new method utilizing supervised machine learning with an
in-domain training set of 1 million tweets built from the hashtags carrying an opinion.

\begin{figure}[!tb]
\centering
\includegraphics[width=0.9\linewidth]{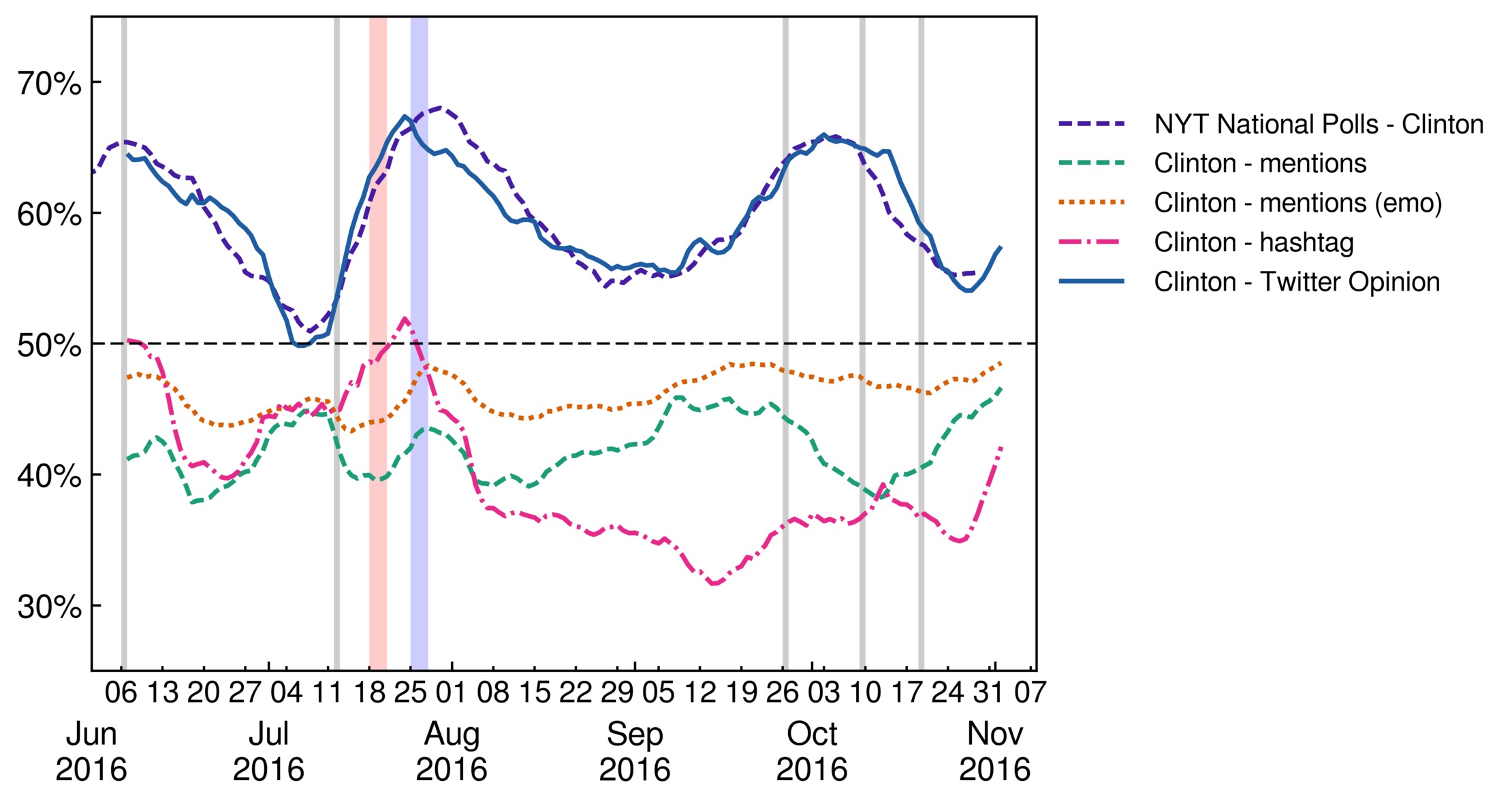}
\caption{\high {\bf Comparison of the fit between different Twitter-based metrics
and the NYT national polling average time series for Hillary Clinton.}
We show the normalized NYT poll scores of Hillary Clinton (dashed purple)
fitted to our Twitter opinion metrics.
The 13 days moving average of Hillary Clinton's score computed using 
mentions (eq. \ref{eq:mentions}, dashed green line),
mentions with sentiment (eq \ref{eq:mentions-sentiment}, dotted orange line), 
hashtags (eq. \ref{eq:hashtags}, dash-dotted pink line)
and our Twitter opinion (continuous blue line).
All the metrics except for our opinion metrics 
are mainly below the 50\% line, and therefore disagree with 
the NYT national polling average.
}
\label{fig:fit_polls_wl13_benchmark}
\end{figure}

}

\section{Discussion}

{\high 
\label{othercontrib}
Using Twitter as a sensor for political opinion has
  attracted enormous attention because there is a general sense that
  digital data may, at some point by the virtue of reaching much
  larger populations, outdate more traditional approaches to polls,
  which is of interest to social science and beyond. Indeed, there
  has been many contributions in
  journals\cite{Tumasjan2011,Borondo2012,Gayo-Avello2013,Caldarelli2014,Kagan2015,Tsakalidis2015,
    Ceron2015,Ceron2016,ceron2016politics,Beauchamp2016} and, in
  social computing
  conferences\cite{OConnor2010,Wang2011,Shi2012,Marchetti-bowick2012,
    Park2012,Hoang2013,Fink2013,
    Thapen2013,Contractor2013,Borge-Holthoefer2015,Saifuddin2016,Wang2016,Llewellyn2016}
  dealing with public opinion and political processes in Twitter (see
  also the Science Issue 6324 (February 03, 2017) on ``Prediction''
  \cite{Bohannon2017,Jasny2017}).

The new method\footnote{available at \url{kcorelab.com}}
we present uses a combination of statistical physics of complex networks,
natural language processing and machine learning
to uncover the opinion of Twitter users and 
to analyze their behavior giving unique insights into the
reason for the observed opinion variations.}
We find a remarkably high correlation between our measured Twitter opinion
trends and the New York Times polling national average.
The opinion trend in Twitter is instantaneous and anticipates the NYT aggregated
surveys by {\shigh 10} days.
{\high We show that using our Twitter opinion trend can be used 
for nowcasting\cite{ceron2016politics} the results of the NYT national polls 7 days in advance with better
accuracy then using a linear extrapolation of the polls, in particular
during abrupt trend reversals.}
This suggests that Twitter can be used as an
early warning signal of global opinion trends happening in the general
population at the country level.

{\high 
\label{discus-bench}
 Our findings demonstrate that measuring the attention
toward candidates does not allow to differentiate the political
support toward each candidate. Indeed, a comparison with previously
proposed methods\cite{OConnor2010,Marchetti-bowick2012,Thapen2013,Jungherr2016}
based on the ratio of users mentioning each candidate show a worse agreement 
with the NYT polling average than our Twitter opinion,
even when sentiment analysis is used to classify
tweets as positive or negative or when hashtags are used to classify users
(see Fig. \FigCompBenchmark).}\\

{\high
\label{discu-poll-failure}
However, the general failure of traditional polls and
  Twitter’s opinion to predict the victory of Donald Trump raises the
  question of whether Twitter can provide insights into this issue not
  accessible to traditional polls. In this regards, we showed the
  necessity of understanding the impact of the difference in activity
  of each supporters group on the final opinion trend to correctly
  interpret it. Our results reveal a difference in the behavior of
  Twitter users supporting Donald Trump and users supporting Hillary
  Clinton. Peaks in the opinion in favor of Clinton are highly
  correlated with large positive fluctuations in the daily number of
  Clinton supporters and coincide with important events such as the
  conventions or the presidential debates. On the other hand, peaks in
  favor of Trump correspond to a lack of mobilization of Clinton
  supporters. Although Clinton supporters are the majority in Twitter,
  Trump supporters are generally more active and more constant in
  their support, while Clinton supporters are less active and show
  their support only occasionally. This dichotomy is also visible in
  the user network dynamics. The strongly connected giant component
  (SCGC), dominated by Trump supporters, shows only small size
  fluctuations and comprises almost only recurring users, as opposed
  to the rest of the network, dominated by Clinton supporters, which
  shows large fluctuations and is where new users arrive. These
  findings confirm previous studies suggesting that right-wing leaning
  Twitter users exhibit greater levels of activity and more tightly
  interconnected social structure \cite{Conover2012,Hoang2013}. We
  push these observations further by showing how these effects
  influence opinion measurement in Twitter. Indeed, our analysis show
  that Twitter’s opinion is mainly measuring the reaction of Clinton
  supporters and not of Trump supporters, suggesting that an important
  part of Trump supporters are missing. This indicates an important
  over-representation in the Twitter population of users expressing their
  opinion toward supporters of the Democratic party. Crucially,
  Twitter’s opinion variations agree remarkably well with the average
  of traditional national polls over a period of more than 5 months
  and consistently precedes it by 10 days. This suggests that the
  demographic imbalance responsible for Twitter’s opinion trend might
  also be present in the traditional polls. In this case, detecting
  such a difference in Twitter supporters behavior is a
  warning for traditional pollsters in the sense that it indicates
  when the polls cannot be trusted as representative of the national 
  elections.
  This is perhaps the most crucial
  question that arose after the 2016 elections. While Twitter may not
  be able to predict the elections, we find that Twitter is able to
  raise clear warning signals that will allow pollsters, politicians
  and the general public in general to know in advance that the
  election day could be a big surprise.}

{\shigh Our results validate the use of our Twitter analytics
machinery as a mean of assessing opinions about political elections
and show the necessity of accompanying the Twitter opinion
with an analysis of user activity in order to correctly
interpret its variations.}
Our analysis comes at a fraction of the cost of traditional NYT
polling methods employed by aggregating the whole of the US\$ 18
billion-revenue market research and public opinion polling industry
(NAICS 54191).
In contrast to traditional unaggregated polling campaigns which are
unscalable, typically ranging at most in the few thousand respondents,
our techniques have the advantage of being highly scalable as they are
only limited by the size of the underlying social networks.
Moreover, traditional polling suffers from a declining rate of respondents being
only 9\% according to current estimates (2012) down from 36\% in
1997\cite{PewRepresentativeness},
while social media is gaining billions of users worldwide.
{\high 
\label{api-bias}
We note that a bias arises from the Twitter's API\cite{Morstatter2013,Gonzalez-Bailon2015}.
Although the demographics representation of Twitter is biased
\cite{PewSocialMedia} and Twitter's API introduces a supplementary unknown
bias in our sample, Twitter allows to study the behavior of
its users and to understand the link between their activity and the
variations in opinion trend, something not accessible to traditional
polls.}

{\high
Provided a large usage of opinion-hashtags and
a polarization of opinion resulting in well separated
hashtag clusters, our approach can be extended to understand other
kind of trend from social media ranging from the opinion 
of users regarding products and brands, to other political movements,
thus, unlocking the power of Twitter to understand trends in the society at large.}\\

\section{Methods}

\subsection{Data collection and social network reconstruction}
\label{sec:net_const}

We continuously collected tweets using the Twitter Search API from
June 1st, 2016 to {\shigh November 8th, 2016}.
We gather a total of {\shigh 171} million
tweets in the English language, mentioning the two top
candidates from the Republican Party (Donald J. Trump) and Democratic
Party (Hillary Clinton) by using two different queries with the
following keywords: \textit{trump OR realdonaldtrump OR donaldtrump}
and \textit{hillary OR clinton OR hillaryclinton}.
{\shigh During this period of 161 days, 15 days are missing
due to connection errors.}
{\high 
\label{filtering}
A more stringent keyword filtering of the dataset
  (see Fig. \FigFilteredResults{} of the Supplementary Information) showed no significant
  changes in our results and conclusions. To asses the importance of
  the possible noise in the data induced by the ``trump'' and
  ``hillary'' keywords, we filtered our dataset to keep only tweets
  with either one of the following keywords : 'realdonaldtrump',
  'hillaryclinton', 'donaldtrump' or at least one of the following
  pairs of keywords 'trump' and 'donald' or 'hillary' and 'clinton'.
  Although this keyword filtering reduces the dataset from 158
  millions tweets to 58 millions tweets (considering only tweets from
  official clients), our results are not significantly changed, as
  shown in Fig. \FigFilteredResults{} in the Supplementary Information, and our
  conclusions still hold.}

For every day in
our dataset, we construct the social network $G(V,E)$ where $V$ is the
set of vertices representing users and $E$ is the set of edges
representing interactions between the users.  In this network, edges
are directed and represent influence.  When a user $v_i \in V$,
retweets, replies to, mentions or quotes any other user $v_j \in V$, a
directed edge is drawn from $v_j$ to $v_i$.  We remove Donald Trump
(\textit{@realdonaldtrump}) and Hillary Clinton
(\textit{@hillaryclinton}) from the network, as we are interested by
the opinion and dynamics of the rest of the network.  We divide the
network in three compartments: the strongly connected giant component
(SCGC), the weakly connected giant component (WCGC) and the corona
(Fig. \ref{fig:GC_CO}).  The SCGC is defined as the largest maximal
set of nodes where there exists a path in both directions between each
pair of nodes. The SCGC is formed by the central, most densely
connected region of the network where the influencers are located, and
where the interactions between users are numerous.  The WCGC is the
largest maximal set of nodes where there exists a path in at least one
direction between each pair of nodes.  The corona is formed by the
smaller components of remaining users and the users that were only
connected to Hillary Clinton or Donald Trump official accounts, which
were removed for consistency.
Users that do not interact with anyone else are not counted in the network,
although we take them into account when computing the opinion of the 
entire dataset (see Fig. \ref{fig:num_users}).

\subsection{Hashtag classification}
\label{sec:ht_class}

{\shigh We split our dataset in two parts.
The first part, from June 1st to September 1st, covers the
two conventions and the second part, from September 1st to November 8th,
covers the three presidential debates until election day.
This allows us to decrease the computational time,
verify the consistency of our results and evaluate the predictive 
quality of our model by training it 
only on the first part of our dataset and evaluating it
on the second part (see Section \ref{sec:fit_nyt}).}

We build a labeled training set of tweets with explicit opinion about
the two presidential candidates by taking advantage of the fact that a
large number of Twitter users label their own tweets by using
hashtags.
The use of a hashtag that explicitly expresses an opinion
in a tweet represents a ``cost'' in terms of self-exposition by
Twitter users\cite{Ceron2015} and therefore allows one to select tweets
that clearly state support or opposition to the candidates.

Our first task is therefore to classify the hashtags
present in our dataset as expressing support or opposition to one of
the candidate.  
{\high For this purpose, we start by identifying the most important
hashtags in term of their total number of occurrences and
then use the relations between hashtags co-occurring in tweets to discover
new  hashtags.

\begin{table}
\centering
 \begin{tabular}{lr}
\toprule
           Hashtag &    Number of occurrences \\
\midrule
          trump &  2240499 \\
      trump2016 &  1320217 \\
           maga &  1139644 \\
        hillary &   905065 \\
 hillaryclinton &   718159 \\
      imwithher &   690519 \\
     trumptrain &   654573 \\
   neverhillary &   634562 \\
   demsinphilly &   627446 \\
     nevertrump &   560876 \\
           tcot &   531389 \\
       rncincle &   498718 \\
   trumppence16 &   473924 \\
    donaldtrump &   409708 \\
 crookedhillary &   396836 \\
\bottomrule
\end{tabular}
\caption{{\high \bf Top occurring hashtags from June 1st to September 1st 2016.}}
\label{tab:tophtgs}
\end{table}

Among the top occurring hashtags (shown in Tab. \ref{tab:tophtgs}),
we identify four hashtags each representing
a different category: \textit{\#maga} for pro-Trump (\textit{maga} is the
abbreviation of the official Trump campaign slogan: {\it Make America Great Again} ,
\textit{\#imwithher} for pro-Hillary (the official Clinton campaign slogan),
\textit{\#nevertrump} for anti-Trump and \textit{\#neverhillary} for anti-Clinton.

We then construct the hashtag co-occurrence network $H(V,E)$, where
the set of vertices ${v_i \in V}$ represents hashtags, and an edge
$e_{ij}$ is drawn between $v_i$ and $v_j$ if they appear together in a
tweet. 
For the period going from June 1st until September 1st, the resulting
graph has 83,159 vertices and 589,566 edges.

Following reference \cite{Martinez-Romo2011}, we test the statistical 
significance of each edge $e_{ij}$ by computing the probability $p_{ij}$ ($p$-value of the
null hypothesis) to observe the corresponding number of co-occurrences by chance knowing 
the number of occurrences $c_i$ and $c_j$ of the vertices $v_i$ and $v_j$, 
and the total number of tweets $N$.
We keep only significant edges satisfying $p<p_0$, where $p_0 = 10^{-6}$,
effectively filtering out spurious relations between hashtags.
Finally, a weight $s_{ij} = \log(p_0/p_{ij})$ representing the significance
of the relation between two hashtags is assigned to each edge.
Retaining only significant edges and considering only the largest component of the filtered graph,
reduces the graph to 8,299 vertices and 26,429 edges.

\label{label-prop}
 Using a method inspired by the method of
label propagation \cite{Raghavan2007},
we use the resulting co-occurrence network to discover hashtags that
are significantly related to the hashtags initially chosen to
represents the different classes.
We simplify the hashtag classification problem by considering 
only two classes: $C_C$ for the hashtags pro-Clinton or anti-Trump and $C_T$ for the
hashtags pro-Trump or anti-Clinton.
Starting from the initial set of hashtags, we infer the class of their 
neighbors $v_i$ by verifying the following condition:
if
\begin{equation}
\sum_{j \in C_C} s_{ij} > \sum_{j \in C_T} s_{ij},
\label{eq:signi1}
\end{equation}
$v_i$ is assigned to $C_C$.
Similarly, if
\begin{equation}
\sum_{j \in C_T} s_{ij} > \sum_{j \in C_C} s_{ij},
\label{eq:signi2}
\end{equation}
$v_i$ is assigned to $C_T$.

We then further filter the new hashtags by keeping only 
hashtags having a number of occurrences
$c_i > r \max\limits_{v_j\in C_k} c_j$ where $c_i$ is the number of 
occurrences of the hashtag associated with vertex $v_i$,
$C_k$ is the class to which $v_i$ belong and $r<1$ is a threshold
parameter that we set to $r=0.001$.
\label{human-valid}
Finally, a human validation among the new hashtags is performed
to only add hashtags that are direct reference to the candidate, its
party or slogans of the candidate and that express an opinion. Table
\TabHashtagsExamples\, shows example of this manual selection.

\begin{table}
\centering
\begin{tabular}{llll}
\toprule
Associated with $C_C$ &  Added to $C_C$
                 &  Associated with $C_T$   &  Added to $C_T$ \\
\midrule
              vote &       strongertogether &        radicalislam &            trumptrain \\
        republican &             donthecon  &                ccot &         trumppence16 \\
           america &               voteblue &          corruption &            votetrump \\
    hillaryclinton &              dumptrump &                ryan &               hillno \\
              real &            hillary2016 &                 fbi &      handcuffhillary \\
            racist &              uniteblue &             hillary &         imnotwithher \\
                p2 &       clintonkaine2016 &                tcot &        votetrump2016 \\
              veep &                hillyes &                jobs &        crookedhillary \\
   trumpuniversity &        nevertrumppence &                tcot &     hillaryforprison \\
               kkk &           chickentrump &              scotus &                 maga \\
\bottomrule
\end{tabular}
\caption{{\high {\bf Example of hashtags discovered in the co-occurrence network.} 
We show hashtags associated with each class but not selected and hashtags selected as they directly 
reference one of the candidate or its party and express an opinion.
The list of hashtags associated with each candidate shows how
the hashtag co-occurrence network can be used discover the topics commonly 
discussed by supporters of each camps.}}
\label{tab:hstg_lists_1st}

\end{table}

This entire process is then repeated adding the newly selected hashtags to each
class and propagating the selection to their neighbors.
After each iteration we also verify the consistency of the classes
by removing hashtags that do no longer satisfy Eqs. (\ref{eq:signi1}) and (\ref{eq:signi2}).

After three iterations of this process, we find a
stable set of hashtags represented in Fig. \ref{fig:ht_cooc}.
\label{comm-detect}
Applying a community detection algorithm\cite{Raghavan2007,Blondel2008}
to the final network found with our method results in four 
different clusters corresponding to the Pro-Clinton, Anti-Clinton, 
Pro-Trump and Anti-Trump hashtags as shown in Fig. \FigHastagsNetwork.
This shows that our resulting classes are well separated and 
correspond to the partition of the network maximizing Newman's modularity.
The full set of hashtags is given in the Supplementary Information 
(Tables \TabHTjunsep{} and \TabHTsepnov).

\label{finalvsinitial}
As a robustness check, we study how the predictions
  using the full set of tweets compares with that using the initial
  seed set of hashtags to train the supervised model. Using the final
  set of hashtags instead of the initial set increases the agreement
  between the Twitter opinion trend and the NYT national polling
  average (see Fig. \FigFinalvsInitial{} of the Supplementary Information). The
  improvement of the classification is also revealed by the larger
  classification scores obtained with the final set of hashtags (see
  Tab. \TabClassificationScores). For example, when using a window length of 13 days, 
  the Pearson correlation coefficient increases 
  from $r = 0.90$ to $r = 0.94$ and the root-mean-square error decreases from RMSE = 0.40\% to RMSE =
0.31\%. The classification improvement in $F_1$-score increases from
$F_1 = 0.73$ to $F_1 = 0.81$.

\label{robust-check}
 To asses the robustness of the manual selection of hashtags, we
perform our daily classification of users using as a basis for our
training set of hashtags three sets, each with a different random
sample containing only 90\% of our final sets of hashtags. We find
that it slightly affect the final daily classification of users
with a root-mean-square deviation of 2.7\% between the ratio of
users in each camp using the reduced sets of hashtags and the full set
of hashtags (see Fig. \FigRobustnessCheck{} of the Supplementary Information). This
indicates that our method is robust against significant (10\%)
variation in the manual selection of hashtags.

}

\subsection{Opinion mining}
\label{sec:opinion_mining}

We build a training set of labeled tweets with two classes: 1)
pro-Clinton or anti-Trump and, 2) pro-Trump or anti-Clinton.  We
discard tweets belonging to the both classes simultaneously to avoid
ambiguous tweets.
We also remove retweets to avoid duplicates in our training set.
We select only tweets that were posted using an
official Twitter client in order to discard tweets that might
originate from bots and to limit the number of tweets posted from
professional accounts.  We use a balanced set, with the same number of
tweets in each class, totaling {\shigh 835,808 tweets for the 
first part of our dataset and 682,508 tweets for the 
second part.}
The tweet contents is
tokenized to extract a list of words, hashtags, usernames, emoticons
and urls.
The hashtags used for labeling are striped off from the tweets and the
other hashtags are kept as they may contain significant information about
the opinion of the tweet.
We also keep the urls as features since they usually 
point to resources determining the opinion of the tweet.
Moreover, replacing all urls by the token ``URL'' (creating an equivalent class)
resulted in smaller classification score.
We use the presence of unigrams and bigrams as features.
{\shigh We find 3.5 million features for the first part of our dataset
and 3.1 million for the second part.}
The performance of different classifiers is tested (Support
Vector Machine, Logistic Regression and modified Huber) with different
regularization methods (Ridge Regression, Lasso and Elastic net)\cite{Hastie2009}.
Hyperparameter optimization is performed with a 10-fold cross
validation optimizing $F_1$ score\cite{scikit-learn}.
The best score is obtained with a
Logistic Regression classifier with $L_2$ penalty (Ridge Regression).
Classification scores are summarized in Table
\ref{tab:cross_val_score}.
{\high We then classify users according to the class in which the
majority of their tweets are classified.

\label{comp_ceron}
In the statistical literature, see for example
  \cite{Hopkins2010,Ceron2016}, it has been shown that, in particular
  when analyzing social media, individual classification error through
  any machine learning approach (as the ones used here) remains high
  and does not vanish due to aggregation because of the large variance
  in estimates. In contrast, it can propagate up to the extent
  that, in many applications with thousands or millions of texts, one
  could see the error increasing to 15\%-20\%.
This could be problematic if one is interested in estimating some type
of aggregate measure through the analysis of social media. In this
respect, instead of performing an individual classification of each
single post and then aggregate the predicted values as done in our
method, algorithms that directly estimates the aggregated distribution
of opinions such as those of \cite{Hopkins2010,Ceron2016} can be more
robust.
Although these methods that directly estimate the aggregated
repartition of opinions are usually more robust and have a higher
accuracy \cite{Hopkins2010,Ceron2016}, they do it at the cost of
losing the individual classification, which is needed for our analysis
of user behavior and supporter partition in the social network.
Moreover, these methods estimate the proportion of documents at the
aggregated level (in our case tweets) in favor of each candidate and
not the proportion of users. As our analysis requires the absolute number 
and the proportion of
users in favor of each candidate, the proportion of the aggregated
tweets will not suffice,
since, typically, users from different parties tweet, in average, at
different rates, see Fig. \FigUsersActivity{}. Therefore, we use our
machine learning methods to directly extract the sentiment of each
individual tweet towards a candidate from where we directly extract
the opinion of each user, which, in turn, can be used to obtain the
percentage of users favoring each candidate.

Using the final set of hashtags instead of the initial set
increases the agreement between the Twitter opinion trend
and the NYT national polling average (see Fig. \FigFinalvsInitial{} of the
Supplementary Information). The improvement of
the classification is also revealed by the 
larger classification scores obtained with 
the final set of hashtags (see Tab.\ref{tab:cross_val_score}).
}

\begin{table}[!tb]
\centering
\begin{tabular}{rlllll}
 \toprule
  & $F_1$ & AUROC & Accuracy & Precision & Recall \\
 \midrule
  Initial set & 0.73 & 0.81 & 0.71 & 0.72 & 0.73\\
  Final set & 0.81 & 0.89 & 0.81 & 0.81 & 0.81 \\
  \bottomrule
\end{tabular}
\caption{\high {\bf Best classification score achieved using a Logistic Regression Classifier with 
$L_2$ regularization.}
For the training set obtained with the final set of hashtags, classification scores are computed over a 10-fold cross-validation.
For the training set obtained with the initial set of hashtags, classification scores are computed on 
the set of tweets contained in the final set but not used for training the classifier.
For $F_1$, Precision and Recall, the average of the two scores computed by taking each class as the 
positive class is computed.}
\label{tab:cross_val_score}
\end{table}

{\high
\subsection{Fit of the Twitter opinion with the National Polling Average}
\label{sec:fit_methods}

Since $r_w^{C}(i) = 1 - r_w^{T}(i)$ where $C$ stand for Clinton and $T$ for Trump
in equation \ref{eq:fit}, we only fit $r_w^{C}(i)$ with its NYT counterpart $y^{C}(i)$.
We have the following relations between the optimal parameters between the Clinton
fit and the Trump fit: $A^{T}$ = $A^{C}$ and $b^{T} = 1 - b^{C} - A^{C}$,
since $y^{C}(i) = 1 - y^{T}(i)$.
We omit the superscript $k$ in the following.

The backward window average at day $i$ of length $w$
days of the support ratio $r$ is defined as

\begin{equation}
\label{eq:bckwrd_ma}
r_w(i) = \frac{1}{w}\sum_{j=i-w+1}^{i} {r}(j)
\end{equation}

where $r(j)$ is the ratio of user in favor of a candidate at day $j$.
The moving window average converts fluctuating daily data into a smooth trend that can be
compared with the NYT smooth time series aggregated over many
polls performed over several days.
We fit the values of $A$, $b$ and $t_d$ for increasing values of $w$
by minimizing the mean squared error. 

Since a backward moving average induces a backward time shift,
the total forward time shift between the Twitter time series and
the NYT polling average is given by $T_d = t_d + \frac{w-1}{2}$,
where we limit $w$ to odd positive integer values to have only integer values.
}

\section{Acknowledgments}

Research was sponsored by Army Research Laboratory under Cooperative
Agreement W911NF-09-2-0053 (ARL Network Science CTA), NSF and NIH.
A. Bovet thanks the Swiss National Science Foundation (SNSF) for the
financial support provided.
The authors thank George Furbish for helping with the data collection.

\section{Author contributions}

H. M., F. M. and A. B. conceived the project.  A. B. performed the analysis
and prepared figures.
A. B. and H. M. wrote the manuscript.

\bibliographystyle{naturemag}
\bibliography{bibliography}

\clearpage

\section*{}

 \begin{center}
 \LARGE{\textbf{Validation of Twitter opinion trends with national polling 
aggregates: Hillary Clinton vs Donald 
Trump - Supplementary Information}}
 
\vspace{1cm}

\large Alexandre Bovet, Flaviano Morone, Hern\'an A. Makse

\vspace{0.2cm} \normalsize \textit{Levich Institute and Physics
  Department, City College of New York, New York, New York 10031, USA}

\end{center}

\begin{figure}[!b]
\centering
\includegraphics[width=0.9\linewidth]{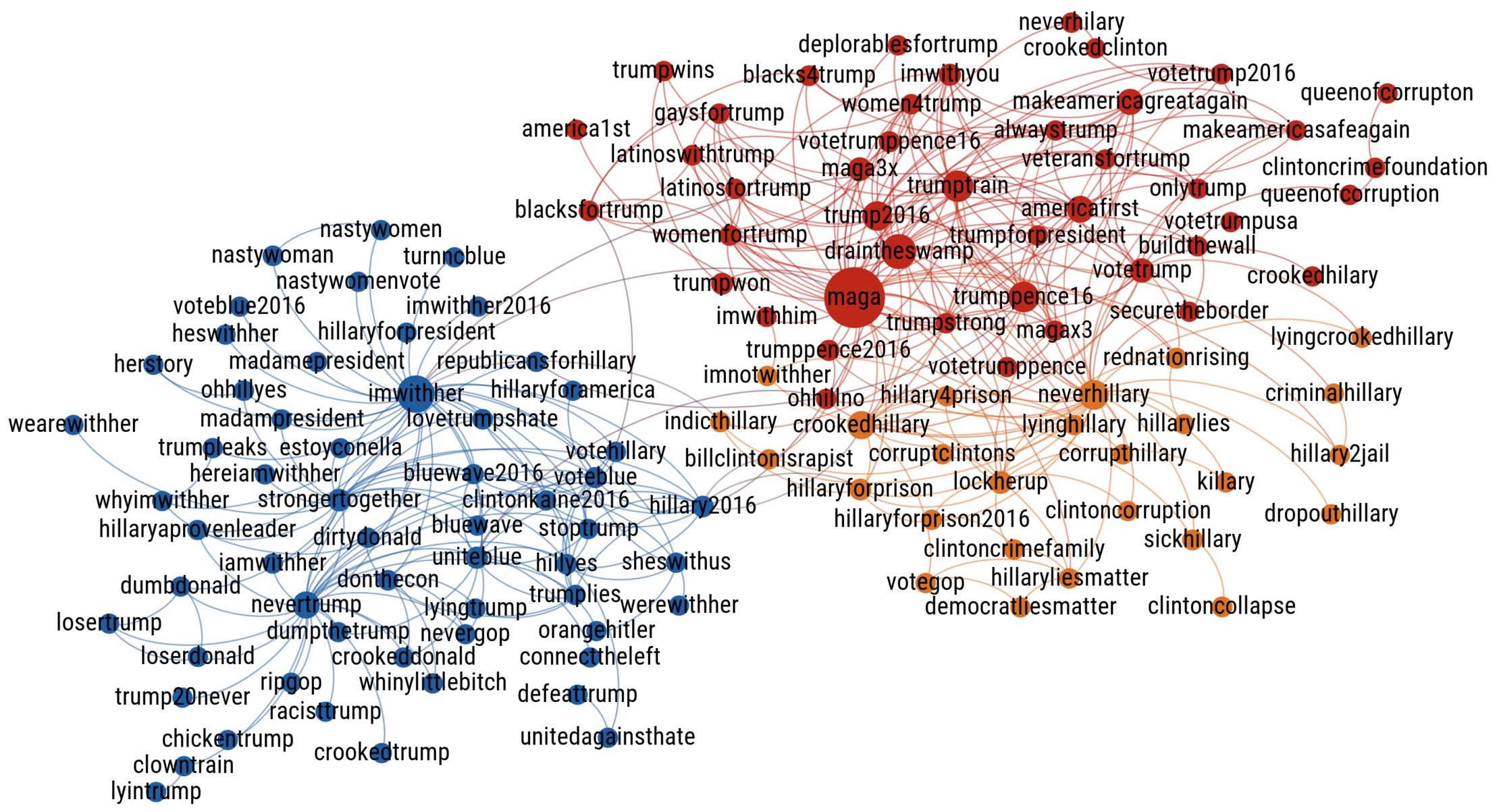}
\caption{{\bf Hashtag classification via network of co-occurrence for September 1st to
November 8th.}
 Network of hashtags obtained by our algorithm from September 1st to November 8th.
  Nodes of the network represent hashtags and an edge is drawn between two hashtags when
  their co-occurrence in tweets is significant.
  The size of the node is proportional to the total number of occurrence of the hashtag.
  Similarly to the network for June 1st to September 1st (Fig. \FigHastagsNetwork{} in the main paper),
  two  main clusters are visible, corresponding to the
  Pro-Trump/Anti-Clinton and Pro-Clinton/Anti-Trump hashtags.
  Inside of these two clusters, the separation between Pro-Trump (red) and
  Anti-Clinton (orange) is visible and the Pro-Clinton and Anti-Trump 
  form a single cluster (blue).
  The coloring corresponds to clusters found by community detection\protect{\cite{Raghavan2007,Blondel2008}}.}
\label{fig:ht_cooc_signi_sep_nov_final_dt}
\end{figure}

\begin{figure}[!tb]
\centering
\includegraphics[width=0.9\linewidth]{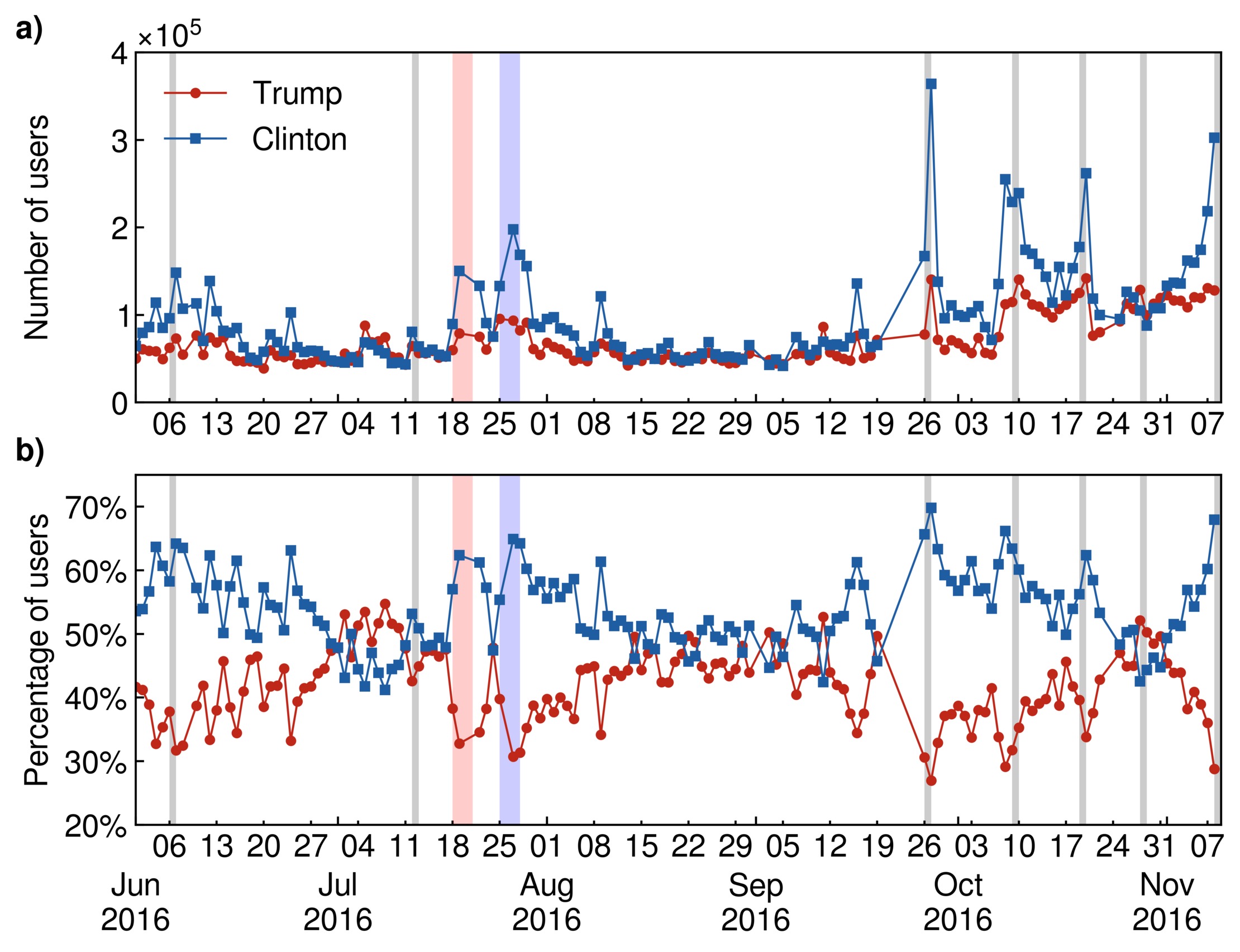}
\caption{{\bf Supporters in the filtered Twitter dataset.}
To asses the importance of the possible noise in the data induced by the 
popular ``trump'' and ``hillary'' keywords, we filter our dataset to keep only
tweets with either one of the following keywords :
'realdonaldtrump', 'hillaryclinton', 'donaldtrump' or
at least one of the following pairs of keywords 'trump' and 'donald' or 'hillary' and 'clinton'. 
Although this keyword filtering reduces the dataset from 158 millions tweets to 58 millions tweets
(considering only tweets from official clients),
our results are not significantly changed.
The relative number of each supporters and the temporal evolution of the number of 
users is similar to the results obtained from our full dataset.}
\label{fig:num_userskwrd_filt}
\end{figure}

\begin{figure}[!tb]
\centering
\includegraphics[width=0.9\linewidth]{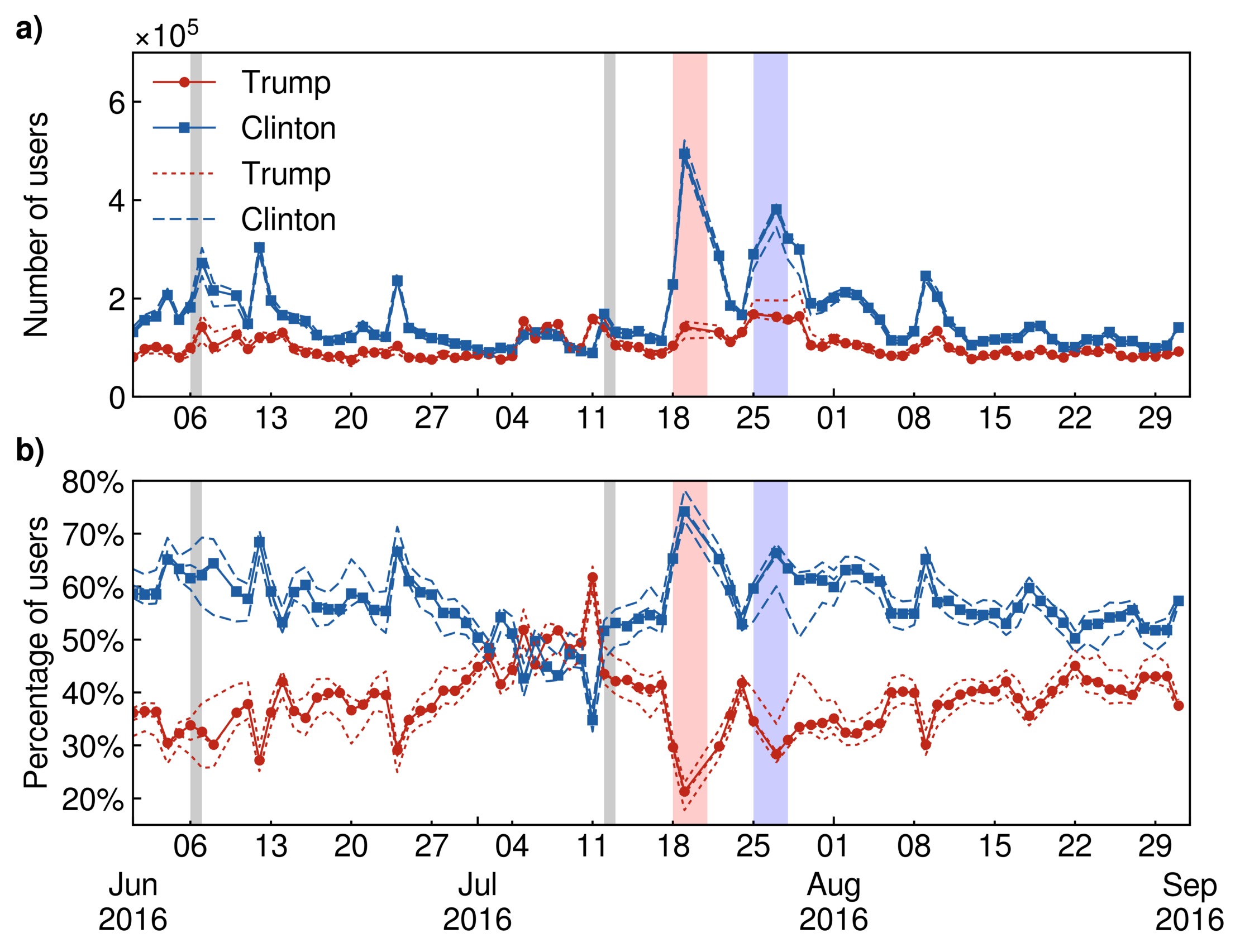}
\caption{{\bf Robustness test of the hashtag selection.}
Comparison of daily users using the full final set of hashtags (continuous lines) with results using a 
random subset with 90\% of the final set of hashtags (dotted and dashed lines) for the
period from June 1st to September 1st.
The root-mean-square deviation between the daily percentage of users found with the full set of hashtags
and with the reduced set is RMSD = 2.7\%.
These results show that our method is robust against variation in the manual section of hashtags used to
build the training set of tweets.}
\label{fig:num_users_random}
\end{figure}

\begin{figure}[!tb]
\centering
\includegraphics[width=0.5\linewidth]{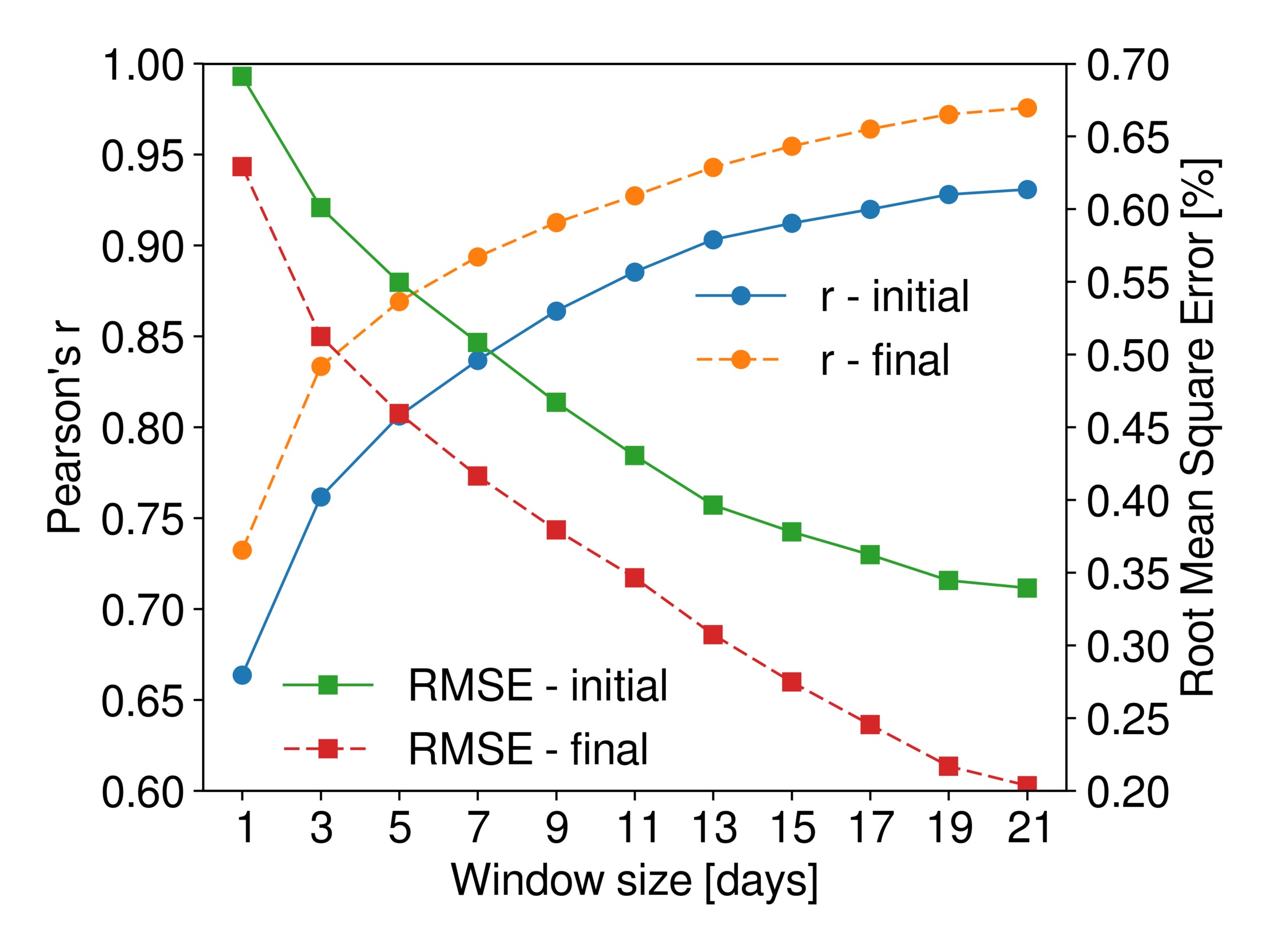}
\caption{{\bf Improvement of the quality of fit when using the final set of hashtags
compared to the initial set.}
Using the final set of hashtags for building our training set instead of the initial set
improve the agreement between our Twitter opinion trend and the NYT Polling Average.
This is shown by a larger Pearson's correlation coefficient ($r$, circles) and a 
lower root-mean-square error (RMSE, squares).
When using a window length of 13 days, the Pearson
correlation coefficient increases from $r = 0.90$ to $r = 0.94$ and
the root-mean-square error decreases from RMSE = 0.40\% to RMSE =
0.31\%.}
\label{fig:fit_pears_rmse_initial_vs_final}
\end{figure}

\begin{figure}[!tb]
\centering
\includegraphics[width=0.75\linewidth]{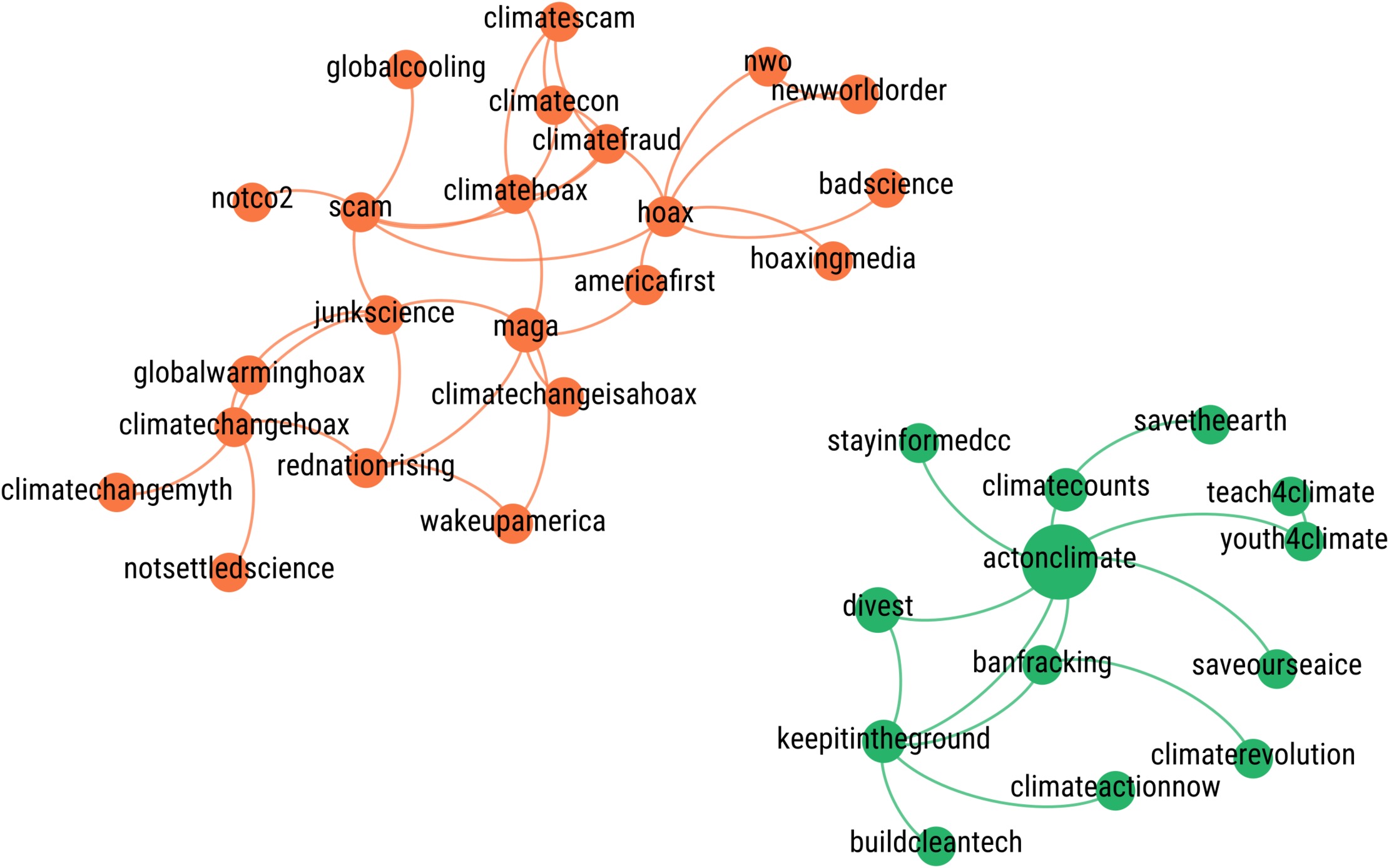}
\caption{{\bf Hashtag co-occurrence networks 
for climate change.}
The network splits into two groups, one with hashtags supporting
action toward climate change (green) and the other with hashtags depicting
climate change as a hoax (orange). This result suggests that our machine
learning and co-occurrence hashtag network method can be generalized
to topics beyond the election setting.
}
\label{fig:ht_cooc_signi_climate_change_final}
\end{figure}

\begin{figure}[!tb]
\centering
\includegraphics[width=0.75\linewidth]{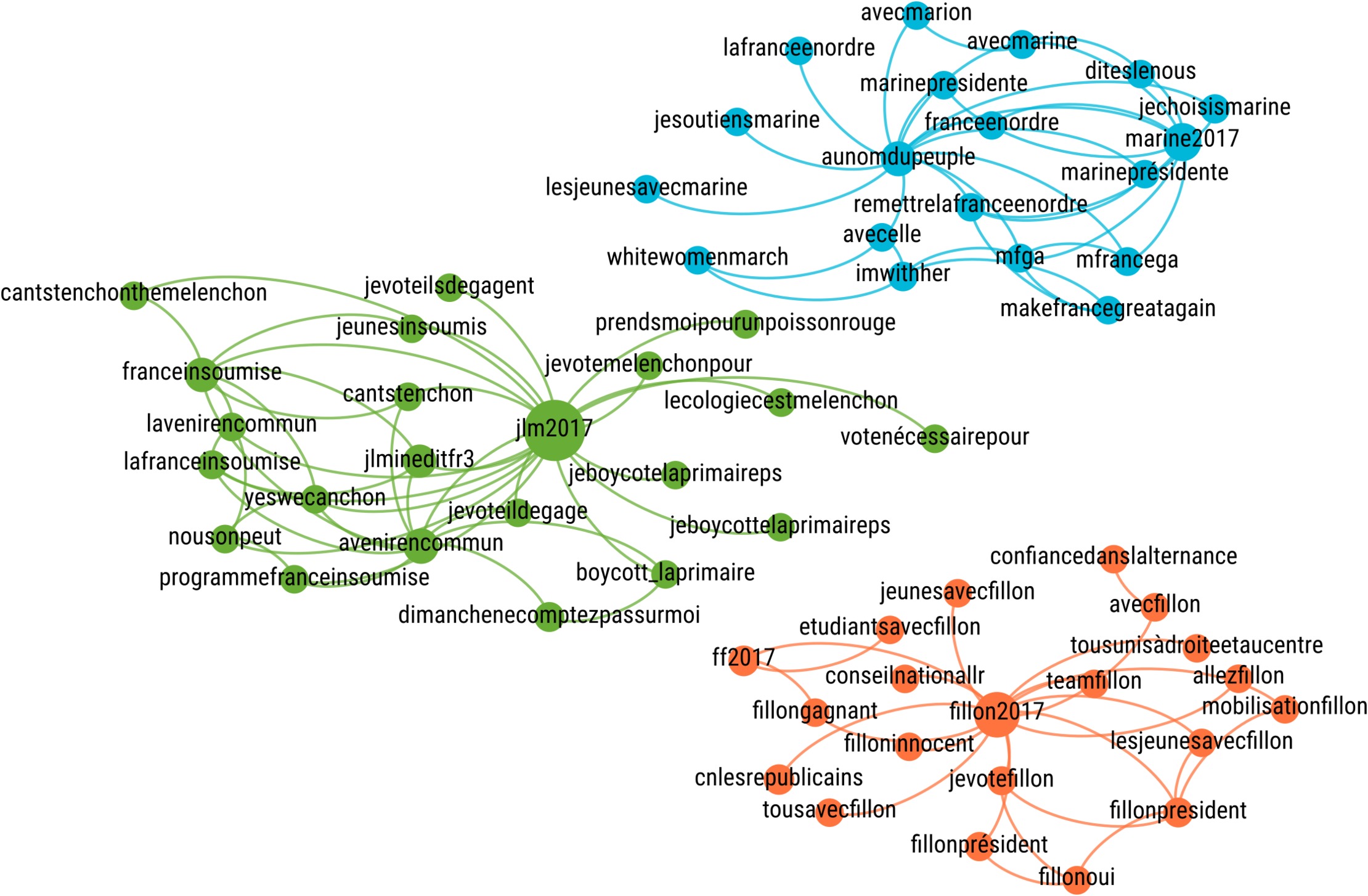}
\caption{{\bf Hashtag co-occurrence networks 
for three candidates of the French presidential elections.}
A separation of the
co-occurrence network into three clear clusters is achieved for the
hashtags employed by users expressing supports to three candidates to
the French presidential election 
(Fran\c{c}ois Fillon (orange), Marine Le Pen (blue) and Jean-Luc M\'elanchon (green)).
They correspond to the three main confirmed candidates at the time of data acquisition
(December 19th, 2016 to January 31st, 2017).
This is the reason why we did not acquire data about Emmanuel Macron, who ultimately
passed the first round of the elections along with Marine Le Pen.
Each group expresses predilection
for each of the three French presidential candidates.
}
\label{fig:ht_cooc_signi_french_elections_final}
\end{figure}
 
\begin{table}[tb]
\centering
\scriptsize
\begin{tabular}{llll}
\toprule
        pro-Clinton &          anti-Trump &             pro-Trump &               anti-Clinton \\
\midrule
           bernwithher &              antitrump &              alwaystrump &             clintoncorruption \\
          bluewave2016 &         anyonebuttrump &            babesfortrump &                  clintoncrime \\
      clintonkaine2016 &           boycotttrump &             bikers4trump &            clintoncrimefamily \\
          estoyconella &           chickentrump &           bikersfortrump &        clintoncrimefoundation \\
              herstory &             clowntrain &             blacks4trump &                corrupthillary \\
            heswithher &          crookeddonald &            buildthatwall &               criminalhillary \\
           hillafornia &          crookeddrumpf &             buildthewall &                crookedclinton \\
           hillary2016 &           crookedtrump &               cafortrump &               crookedclintons \\
     hillaryforamerica &           crybabytrump &          democrats4trump &                 crookedhilary \\
          hillaryforpr &            defeattrump &  donaldtrumpforpresident &                crookedhiliary \\
   hillaryforpresident &            dirtydonald &             feelthetrump &                crookedhillary \\
 hillarysopresidential &              donthecon &    feminineamerica4trump &         crookedhillaryclinton \\
    hillarysoqualified &                 drumpf &               gays4trump &                 deletehillary \\
         hillarystrong &             dumbdonald &             gaysfortrump &                dropouthillary \\
         hillstorm2016 &           dumpthetrump &                  gotrump &                   fbimwithher \\
               hillyes &              dumptrump &                heswithus &               handcuffhillary \\
               hrc2016 &       freethedelegates &                imwithhim &              heartlesshillary \\
       hrcisournominee &    lgbthatestrumpparty &                imwithyou &                  hillary2jail \\
            iamwithher &            loserdonald &            latinos4trump &                  hillary4jail \\
              imwither &             losertrump &          latinosfortrump &                hillary4prison \\
             imwithher &         lovetrumpshate &                     maga &            hillary4prison2016 \\
         imwithher2016 &        lovetrumpshates &         makeamericagreat &              hillaryforprison \\
         imwithhillary &             lyindonald &    makeamericagreatagain &          hillaryforprison2016 \\
            imwiththem &            lyingdonald &     makeamericasafeagain &         hillaryliedpeopledied \\
             itrusther &             lyingtrump &     makeamericaworkagain &                   hillarylies \\
         itrusthillary &              lyintrump &                onlytrump &             hillaryliesmatter \\
       madamepresident &  makedonalddrumpfagain &           presidenttrump &                 hillarylostme \\
        madampresident &               nevergop &          rednationrising &          hillaryrottenclinton \\
     momsdemandhillary &             nevertrump &                  trump16 &              hillarysolympics \\
             ohhillyes &        nevertrumppence &                trump2016 &                        hillno \\
       readyforhillary &          nodonaldtrump &               trumpcares &              hypocritehillary \\
   republicans4hillary &                notrump &        trumpforpresident &                  imnotwithher \\
 republicansforhillary &         notrumpanytime &           trumpiswithyou &                 indicthillary \\
            sheswithus &             poordonald &             trumppence16 &        iwillneverstandwithher \\
   standwithmadampotus &            racisttrump &           trumppence2016 &                       killary \\
      strongertogether &      releasethereturns &              trumpstrong &                     lockherup \\
             uniteblue &       releaseyourtaxes &               trumptrain &           lyingcrookedhillary \\
          vote4hillary &                 ripgop &         veteransfortrump &                  lyinghillary \\
              voteblue &        showusyourtaxes &               vets4trump &                   lyinhillary \\
          voteblue2016 &           sleazydonald &                  votegop &        moretrustedthanhillary \\
           votehillary &              stoptrump &                votetrump &                  neverclinton \\
         welovehillary &            stupidtrump &            votetrump2016 &              nevereverhillary \\
            yeswekaine &           traitortrump &       votetrumppence2016 &                  neverhillary \\
                       &        treasonoustrump &              woman4trump &                 neverhilllary \\
                       &           trump20never &              women4trump &                 nohillary2016 \\
                       &              trumplies &            womenfortrump &                nomoreclintons \\
                       &        trumpliesmatter &                          &                    notwithher \\
                       &            trumpsopoor &                          &                      ohhillno \\
                       &          trumpthefraud &                          &         releasethetranscripts \\
                       &        trumptrainwreck &                          &                  riskyhillary \\
                       &           trumptreason &                          &                       shelies \\
                       &             unfittrump &                          &                   sickhillary \\
                       &             weakdonald &                          &                   stophillary \\
                       &      wherertrumpstaxes &                          &               stophillary2016 \\
                       &        wheresyourtaxes &                          &       theclintoncontamination \\
                       &       whinylittlebitch &                          &                 wehatehillary \\
                       &       womentrumpdonald &                          &  whatmakeshillaryshortcircuit \\
\bottomrule
\end{tabular}
  \caption{{\bf List of hashtags used for labeling the tweet training set from June 1st to September 1st.}}
  \label{tab:hashtags_final_set}
\end{table}

\begin{table}[tb]
\centering
\scriptsize
\begin{tabular}{llll}
\toprule
        pro-Clinton &          anti-Trump &             pro-Trump &               anti-Clinton \\
\midrule
\midrule
              bluewave &      chickentrump &            alwaystrump &     billclintonisrapist \\
          bluewave2016 &        clowntrain &             america1st &         clintoncollapse \\
      clintonkaine2016 &     crookeddonald &           americafirst &       clintoncorruption \\
        connecttheleft &      crookedtrump &           blacks4trump &      clintoncrimefamily \\
          estoyconella &       defeattrump &         blacksfortrump &  clintoncrimefoundation \\
        hereiamwithher &       dirtydonald &           buildthewall &         corruptclintons \\
              herstory &         donthecon &    deplorablesfortrump &          corrupthillary \\
            heswithher &        dumbdonald &          draintheswamp &         criminalhillary \\
           hillary2016 &      dumpthetrump &           gaysfortrump &          crookedclinton \\
  hillaryaprovenleader &       loserdonald &              imwithhim &           crookedhilary \\
     hillaryforamerica &        losertrump &              imwithyou &          crookedhillary \\
   hillaryforpresident &    lovetrumpshate &        latinosfortrump &      democratliesmatter \\
               hillyes &        lyingtrump &       latinoswithtrump &          dropouthillary \\
            iamwithher &         lyintrump &                   maga &            hillary2jail \\
             imwithher &        nastywoman &                 maga3x &          hillary4prison \\
         imwithher2016 &        nastywomen &                 magax3 &        hillaryforprison \\
       madamepresident &    nastywomenvote &  makeamericagreatagain &    hillaryforprison2016 \\
        madampresident &          nevergop &   makeamericasafeagain &             hillarylies \\
             ohhillyes &        nevertrump &              onlytrump &       hillaryliesmatter \\
 republicansforhillary &      orangehitler &        rednationrising &            imnotwithher \\
            sheswithus &       racisttrump &        securetheborder &           indicthillary \\
      strongertogether &            ripgop &              trump2016 &                 killary \\
            turnncblue &         stoptrump &      trumpforpresident &               lockherup \\
             uniteblue &      trump20never &           trumppence16 &     lyingcrookedhillary \\
     unitedagainsthate &        trumpleaks &         trumppence2016 &            lyinghillary \\
              voteblue &         trumplies &            trumpstrong &             neverhilary \\
          voteblue2016 &  whinylittlebitch &             trumptrain &            neverhillary \\
           votehillary &                   &              trumpwins &                ohhillno \\
          wearewithher &                   &               trumpwon &       queenofcorruption \\
           werewithher &                   &       veteransfortrump &        queenofcorrupton \\
          whyimwithher &                   &                votegop &             sickhillary \\
                       &                   &              votetrump &                         \\
                       &                   &          votetrump2016 &                         \\
                       &                   &         votetrumppence &                         \\
                       &                   &       votetrumppence16 &                         \\
                       &                   &           votetrumpusa &                         \\
                       &                   &            women4trump &                         \\
                       &                   &          womenfortrump &                         \\
\bottomrule
\end{tabular}
  \caption{{\bf List of hashtags used for labeling the tweet training set from September 1st to November 8th.}}
  \label{tab:hashtags_final_set_sep_nov}
\end{table}

\end{document}